\documentclass[aps,pra,reprint,showpacs]{revtex4-1}
\usepackage[dvipdfm]{graphicx}
\usepackage{amsmath,amssymb}
\usepackage{bm}
\usepackage{ulem}
\begin{document}

%\preprint{}

\title{Metastability, excitations, fluctuations, and multiple-swallowtail structures of a superfluid in a Bose-Einstein condensate in the presence of a uniformly moving defect}

\author{Masaya Kunimi}
\thanks{Present address : Department of Engineering Science, University of Electro-Communications, Tokyo 182-8585, Japan}
%\email{E-mail : kunimi@vortex.c.u-tokyo.ac.jp}
\email{E-mail : kunimi@hs.pc.uec.ac.jp}
\affiliation{Department of Basic Science, The University of Tokyo, Tokyo 153-8902, Japan}
\author{Yusuke Kato}
\affiliation{Department of Basic Science, The University of Tokyo, Tokyo 153-8902, Japan}

%\thanks{}

\date{\today}

%%%%%%%%%%%%%%%%%%%%%%%%%%%%%%%%%%%%%%%%%%%%%%%%%%%%
\begin{abstract}
We solve the Gross-Pitaevskii (GP) and Bogoliubov equations to investigate the metastability of superfluidity in a Bose-Einstein condensate in the presence of a uniformly moving defect potential in a two-dimensional torus. We calculate the total energy and momentum as functions of the driving velocity of the moving defect and find metastable states with negative effective-mass near the critical velocity. We also find that the first excited energy (energy gap) in the finite-sized torus closes at the critical velocity, that it obeys one-fourth power-law scaling, and that the dynamical fluctuation of the density (amplitude of the order parameter) is strongly enhanced near the critical velocity. We confirm the validity of our results near the critical velocity by calculating the quantum depletion. We find an unconventional swallowtail structure (multiple-swallowtail structure) through calculations of the unstable stationary solutions of the GP equation. 
\end{abstract}
%%%%%%%%%%%%%%%%%%%%%%%%%%%%%%%%%%%%%%%%%%%%%%%%%%%%

\pacs{67.85.De, 03.75.Lm, 03.75.Kk}
%\keywords{}
\maketitle

%%%%%%%%%%%%%%%%%%%%%%%%%%%%%%%%%%%%%%%%%%%%%%%%%%%%%%%%%%%%%%
\section{Introduction}\label{sec:Introduction}
%%%%%%%%%%%%%%%%%%%%%%%%%%%%%%%%%%%%%%%%%%%%%%%%%%%%%%%%%%%%%%

The breakdown of superfluidity is a long-standing but still central issue regarding quantum fluids \cite{Landau1941,Feynman1955,Leggett1973,Nozieres1990,Leggett2001,Leggett2006}. It has been observed in experiments of cold atomic gases trapped in simply connected  geometry \cite{Raman1999,Onofrio2000,Inouye2001,Engels2007,Neely2010,Desbuquois2012} and multiply connected geometry \cite{Ryu2007,Ramanathan2011,Moulder2012,Beattie2013,Neely2013,Ryu2013,Wright2013,Wright2013_full,Ryu2014,Eckel2014,Jendrzejewski2014,Eckel2014_2,Corman2014}. The latter experiments have exhibited various properties of superfluidity, including critical velocity, vortex nucleation, decay of persistent current, phase slip, and hysteresis. The breakdown of superflow stability in cold atoms has been studied theoretically \cite{Frisch1992,Pomeau1993,Hakim1997,Jackson1998,Josserand1999,Winiecki1999,Jackson2000,Huepe2000,Brand2001,Pham2002,Pavloff2002,Aftalion2003,El2006,Piazza2009,Sasaki2010,Aioi2011,Woo2012,Dubessy2012,Piazza2013} using the Gross-Pitaevskii (GP) equation \cite{Gross1961,Pitaevskii1961}. For example, Frisch {\it et al}. \cite{Frisch1992} showed that, in the presence of a defect potential, a vortex pair is nucleated when the velocity is above a critical value. Nucleation of solitons was studied in one-dimensional superfluids \cite{Hakim1997}, and the results aid in our understanding of the relationship between the nucleation of a topological defect and the breakdown of superfluidity. 

The breakdown of superfluidity can be understood, within the mean-field theory, through an energy diagram of stable or unstable states as functions of a control parameter (angular velocity of a container or driving velocity of an optical lattice, for example). Superfluidity breaks down at the control parameter when the metastable superflow state meets an unstable state in the energy diagram. From condensate wave function of the unstable state, furthermore, we see that dynamics of topological defects causes decay of superflow.  
 
As a typical energy diagram of superfluids, swallowtail structure \cite{Diakonov2002,Mueller2002,Machholm2003,Machholm2004,Wu2003,Menotti2003_2,Taylor2003,Seaman2005,Morsch2006,Danshita2007,Kanamoto2009,Fialko2012,Baharian2013} has been investigated for one-dimensional optical lattices and ring-shaped systems with narrow widths. Those theories seem to explain experimental results. More recently, experimental results \cite{Wright2013,Wright2013_full} on vortex nucleations and breakdown of superfluids have been discussed on the basis of swallowtail structure and corresponding energy landscape.  

In two- and three-dimensional superfluids, however, the whole structure of energy diagram of stable or unstable branches has not yet been known. Our aim is to find the whole structure of energy diagram in two-dimensional systems and gain physical insight into breakdown of superfluids related to vortex nucleation. For this purpose, we solve the GP and Bogoliubov equations in two-dimensional torus with a uniformly moving defect potential. In our previous work \cite{Kunimi2014}, we showed the properties of excitations and fluctuations near the critical velocity. In this full paper, we present the whole structure of energy diagram and related results; material not reported in Ref.~\cite{Kunimi2014} includes the existence or absence of a ghost vortex pair, quantum depletion near the critical velocity, and energy diagram, which we call {\it multiple-swallowtail structure}.

This paper is organized as follows: In Sec.~\ref{sec:Model}, we introduce our model. In Sec.~\ref{sec:Results}, we present the stable stationary solutions of the GP and Bogoliubov equations \cite{Bogoliubov1947,Fetter1972}. In Sec.~\ref{subsec:energy-diagram_2D}, the velocity dependence of the total energy and the total momentum are presented, and we compare our results with those for optical lattice systems. In Sec.~\ref{subsec:density_and_phase_profile}, we show the density and the phase profiles below and above the critical velocity and discuss the appearance of a ghost vortex pair. In Secs.~\ref{subsec:excitation_spectra_2D} and \ref{subsec:fluctuation_2D}, we demonstrate the properties of the excitation and the fluctuation. In Sec.~\ref{subsec:Quantum depletion}, we show the results for quantum depletion and discuss the validity of the GP and the Bogoliubov approximation in this system. In Sec.~\ref{sec:Results_unstable}, we present unstable stationary solutions of the GP equation. We show that a multiple-swallowtail structure appears in this system. In Sec.~\ref{sec:discussion}, we discuss the bifurcation structure of the system, the relation between the fluctuations and the energy landscape near the critical velocity, and the possible effects of the multiple-swallowtail structure on the decay of supercurrent. Finally, we summarize our results in Sec.~\ref{sec:Summary}. The numerical methods used in the present work are summarized in Appendixes \ref{app:Methods_of_numerical_calculations} and \ref{app:PACM}.

%%%%%%%%%%%%%%%%%%%%%%%%%%%%%%%%%%%%%%%%%%%%%%%%%%%%%%%%%%%%%%
\section{Model}\label{sec:Model}
%%%%%%%%%%%%%%%%%%%%%%%%%%%%%%%%%%%%%%%%%%%%%%%%%%%%%%%%%%%%%%
%%%%%%%%%%%%%%%%%%%%%%%%%%%%%%%%%%%%%%%%%%%%%%%%%%%%%%%%%%%%%%
\subsection{GP equation in a laboratory frame}\label{subsec:GP_equation_in_laboratory_frame}
%%%%%%%%%%%%%%%%%%%%%%%%%%%%%%%%%%%%%%%%%%%%%%%%%%%%%%%%%%%%%%

We consider a system in which $N$ bosons of mass $m$ are confined in a two-dimensional torus $[-L/2, +L/2)\times[-L/2, +L/2)$. In the mean-field approximation, the physical properties of the system can be described by a complex order parameter (condensate wave function) $\Psi_{\rm L}(\bm{r}_{\rm L}, t_{\rm L})$, where $\bm{r}_{\rm L}$ and $t_{\rm L}$ denote the coordinate and time in the laboratory frame, respectively. The subscript ${\rm L}$ denotes the variables in the laboratory frame. The condensate wave function obeys the GP equation \cite{Gross1961,Pitaevskii1961}:
\begin{align}
i\hbar\frac{\partial}{\partial t_{\rm L}}\Psi_{\rm L}(\bm{r}_{\rm L}, t_{\rm L})=&-\frac{\hbar^2}{2m}{\bm \nabla}_{\rm L}^2\Psi_{\rm L}(\bm{r}_{\rm L}, t_{\rm L})\nonumber \\
&+U(\bm{r}_{\rm L}+\bm{v}t_{\rm L})\Psi_{\rm L}(\bm{r}_{\rm L}, t_{\rm L})\nonumber \\
&\quad +g|\Psi_{\rm L}(\bm{r}_{\rm L}, t_{\rm L})|^2\Psi_{\rm L}(\bm{r}_{\rm L}, t_{\rm L}),\label{eq:Model:GP_equation_in_lab_frame}
\end{align}
where $U(\bm{r}_{\rm L}+\bm{v}t_{\rm L})$ represents a moving defect potential with a constant velocity $-\bm{v}$ and $g(>0)$ is the strength of the interaction. We use the Gaussian potential:
\begin{eqnarray}
U(\bm{r}_{\rm L})\equiv U_0\exp{\left[-\left(\frac{\bm{r}_{\rm L}}{d}\right)^2\right]},\label{eq:Model_defect_potential}
\end{eqnarray}
where $U_0(>0)$ and $d$ are the strength and the width of the potential, respectively. Throughout this paper, the velocity of the potential is in the direction of positive $x$ ($\bm{v}\equiv v\bm{e}_x$, where $\bm{e}_x$ is a unit vector in the direction of $x$.) Periodic boundary conditions are imposed on $\Psi_{\rm L}(\bm{r}_{\rm L}, t_{\rm L})$, because there is a requirement that the condensate wave function should be single valued:
\begin{eqnarray}
\Psi_{\rm L}(\bm{r}_{\rm L}+L\bm{e}_x, t_{\rm L})&=&\Psi_{\rm L}(\bm{r}_{\rm L}, t_{\rm L}),\label{eq:Model_periodic_boundary_conditin_x}\\
\Psi_{\rm L}(\bm{r}_{\rm L}+L\bm{e}_y, t_{\rm L})&=&\Psi_{\rm L}(\bm{r}_{\rm L}, t_{\rm L}),\label{eq:Model_periodic_boundary_conditin_y}
\end{eqnarray}
where $\bm{e}_y$ is a unit vector in the direction of positive $y$. From this boundary condition, we can define the winding number:
\begin{eqnarray}
W\equiv \frac{1}{2\pi}\int^{+L/2}_{-L/2}dx_{\rm L}\frac{\partial}{\partial x_{\rm L}}\varphi_{\rm L}(\bm{r}_{\rm L}, t_{\rm L}),\label{eq:Model_definition_of_winding_number}
\end{eqnarray}
where $\varphi_{\rm L}(\bm{r}_{\rm L}, t_{\rm L})$ is the phase of the condensate wave function.

%%%%%%%%%%%%%%%%%%%%%%%%%%%%%%%%%%%%%%%%%%%%%%%%%%%%%%%%%%%%%%
\subsection{GP equation in a moving frame}\label{subsec:GP_equation_in_moving_frame}
%%%%%%%%%%%%%%%%%%%%%%%%%%%%%%%%%%%%%%%%%%%%%%%%%%%%%%%%%%%%%%
The GP equation in the laboratory frame (\ref{eq:Model:GP_equation_in_lab_frame}) depends explicitly on time. We remove $t$ dependence by performing a coordinate transformation from the laboratory frame to a moving frame \cite{Leggett1973,Leggett2006}, as follows:
\begin{align}
\bm{r}&\equiv \bm{r}_{\rm L}+\bm{v}t_{\rm L},\label{eq:Model_change_of_the_coordinate}\\
t&\equiv t_{\rm L},\label{eq:Model_change_of_the_time}\\
\Psi(\bm{r}, t)&\equiv \exp{\left(\frac{i}{\hbar}\frac{1}{2}m\bm{v}^2t_{\rm L}+\frac{i}{\hbar}m\bm{v}\cdot\bm{r}_{\rm L}\right)}\Psi_{\rm L}(\bm{r}_{\rm L}, t_{\rm L}),\label{eq:Model_local_gauge_transoformation}\\
{\bm \nabla}_{\rm L}&={\bm \nabla},\label{eq:Model_transformation_nabla}\\
\frac{\partial}{\partial t_{\rm L}}&=\frac{\partial}{\partial t}+\bm{v}\cdot{\bm \nabla}.\label{eq:Model_transformation_time_derivative}
\end{align}
Using Eqs.~(\ref{eq:Model_change_of_the_coordinate})$-$(\ref{eq:Model_transformation_time_derivative}), the GP equation in the moving frame is given by
\begin{align}
i\hbar\frac{\partial}{\partial t}\Psi(\bm{r}, t)&=-\frac{\hbar^2}{2m}\nabla^2\Psi(\bm{r}, t)+U(\bm{r})\Psi(\bm{r}, t)\nonumber \\
&\quad +g|\Psi(\bm{r}, t)|^2\Psi(\bm{r}, t).\label{eq:Model_GP_equation_in_mov_frame}
\end{align}
As a result of transformation (\ref{eq:Model_local_gauge_transoformation}), the periodic boundary condition becomes twisted \cite{Lieb2002_2}:
\begin{eqnarray}
\Psi(\bm{r}+L\bm{e}_x, t)&=&e^{imvL/\hbar}\Psi(\bm{r}, t),\label{eq:Model_twisted_periodic_boundary_condtion_x}\\
\Psi(\bm{r}+L\bm{e}_y, t)&=&\phantom{e^{imvL/\hbar}}\Psi(\bm{r}, t).\label{eq:Model_twisted_periodic_boundary_condtion_y}
\end{eqnarray}

The stationary solution of the GP equation (\ref{eq:Model_GP_equation_in_mov_frame}) is given by $\Psi(\bm{r}, t)=e^{-i\mu t/\hbar}\Psi(\bm{r})$, where $\mu$ is the chemical potential. Substituting this relation into Eq.~(\ref{eq:Model_GP_equation_in_mov_frame}), we obtain the time-independent GP equation:
\begin{align}
-\frac{\hbar^2}{2m}\nabla^2\Psi(\bm{r})+U(\bm{r})\Psi(\bm{r})+g|\Psi(\bm{r})|^2\Psi(\bm{r})=\mu\Psi(\bm{r}).\label{eq:Model_time-independent_GP_equation_in_mov_frame}
\end{align}
The chemical potential $\mu$ is determined by the condition
\begin{eqnarray}
N=\int d\bm{r}|\Psi(\bm{r})|^2.\label{eq:Model_total_particle_number_condtion}
\end{eqnarray}

%%%%%%%%%%%%%%%%%%%%%%%%%%%%%%%%%%%%%%%%%%%%%%%%%%%%%%%%%%%%%%
\subsection{Bogoliubov equation}\label{subsec:Bogoliubov_equation}
%%%%%%%%%%%%%%%%%%%%%%%%%%%%%%%%%%%%%%%%%%%%%%%%%%%%%%%%%%%%%%
The Bogoliubov equation \cite{Bogoliubov1947,Fetter1972} can be derived by linearizing the GP equation around the stationary solution $\Psi(\bm{r})$. Substituting
\begin{align}
\Psi(\bm{r}, t)\equiv e^{-i\mu t/\hbar}\left[\Psi(\bm{r})+u_i(\bm{r})e^{-i\epsilon_i t/\hbar}-v^{\ast}_i(\bm{r})e^{i\epsilon_i^{\ast}t/\hbar}\right]\label{eq:Model_linearizing_GP_equation}
\end{align}
into the time-dependent GP equation (\ref{eq:Model_GP_equation_in_mov_frame}), and neglecting the higher-order terms of $u_i(\bm{r})$ and $v_i(\bm{r})$, we obtain the Bogoliubov equation,
\begin{align}
\begin{bmatrix}
\mathcal{L} & -g[\Psi(\bm{r})]^2 \\
g[\Psi^{\ast}(\bm{r})]^2 & -\mathcal{L}
\end{bmatrix}
\begin{bmatrix}
u_i(\bm{r}) \\
v_i(\bm{r})
\end{bmatrix}
=\epsilon_i
\begin{bmatrix}
u_i(\bm{r}) \\
v_i(\bm{r})
\end{bmatrix}
,\label{eq:Model_Bogoliubov_equation}\\
\mathcal{L}\equiv -\frac{\hbar^2}{2m}\nabla^2+U(\bm{r})-\mu+2g|\Psi(\bm{r})|^2.\label{eq:Model_definition_of_L}
\end{align}
Here $u_i(\bm{r})$ and $v_i(\bm{r})$ are the wave functions of the $i$-th excited state with an excitation energy $\epsilon_i$. The boundary conditions for $u_i(\bm{r})$ and $v_i(\bm{r})$ are determined by the condition in which Eq.~(\ref{eq:Model_linearizing_GP_equation}) satisfies the twisted periodic boundary conditions (\ref{eq:Model_twisted_periodic_boundary_condtion_x}) and (\ref{eq:Model_twisted_periodic_boundary_condtion_y}):
\begin{eqnarray}
u_i(\bm{r}+L\bm{e}_x)&=&e^{+i m v L/\hbar}u_i(\bm{r}),\label{eq:Model_boundary_condtion_for_u_x}\\
v_i(\bm{r}+L\bm{e}_x)&=&e^{-i m v L/\hbar}v_i(\bm{r}),\label{eq:Model_boundary_condtion_for_v_x}\\
u_i(\bm{r}+L\bm{e}_y)&=&\phantom{e^{+i m v L/\hbar}}u_i(\bm{r}),\label{eq:Model_boundary_condtion_for_u_y}\\
v_i(\bm{r}+L\bm{e}_y)&=&\phantom{e^{-i m v L/\hbar}}v_i(\bm{r}).\label{eq:Model_boundary_condtion_for_v_y}
\end{eqnarray}
The wave functions for the excited states satisfy the following orthonormal conditions:
\begin{eqnarray}
\int d\bm{r}\left[u_i^{\ast}(\bm{r})u_j(\bm{r})-v_i^{\ast}(\bm{r})v_j(\bm{r})\right]&=&\delta_{i j},\label{eq:Model_orthonormal_condition_for_u_and_v_1}\\
\int d\bm{r}\left[u_i(\bm{r})v_j(\bm{r})-v_i(\bm{r})u_j(\bm{r})\right]&=&0.\label{eq:Model_orthonormal_condition_for_u_and_v_2}
\end{eqnarray}

Throughout this paper, length, energy, and time are normalized by the healing length $\xi\equiv \hbar/\sqrt{m g n_0}$, $\epsilon_0\equiv g n_0$, and $\tau\equiv \hbar/\epsilon_0$, respectively, where $n_0\equiv N/S$ ($S\equiv L^2$ is the area of the system) is the mean particle density. The velocity is normalized by the sound velocity $v_{\rm s}\equiv \sqrt{g n_0/m}$ or $v_0\equiv 2\pi\hbar/(m L)$.

Numerically solving the GP and Bogoliubov equations yields the condensate wave function, excitation spectra, and  wave functions for the excited states. The methods we used for the numerical calculations are summarized in Appendixes \ref{app:Methods_of_numerical_calculations} and \ref{app:PACM}.

%%%%%%%%%%%%%%%%%%%%%%%%%%%%%%%%%%%%%%%%%%%%%%%%%%%%%%%%%%%%%%
\section{Results for Stable Branches}\label{sec:Results}
%%%%%%%%%%%%%%%%%%%%%%%%%%%%%%%%%%%%%%%%%%%%%%%%%%%%%%%%%%%%%%

%%%%%%%%%%%%%%%%%%%%%%%%%%%%%%%%%%%%%%%%%%%%%%%%%%%%%%%%%%%%%%
\subsection{Energy and momentum}\label{subsec:energy-diagram_2D}
%%%%%%%%%%%%%%%%%%%%%%%%%%%%%%%%%%%%%%%%%%%%%%%%%%%%%%%%%%%%%%

\begin{figure}[t]
\centering
\includegraphics[width=8.0cm,clip]{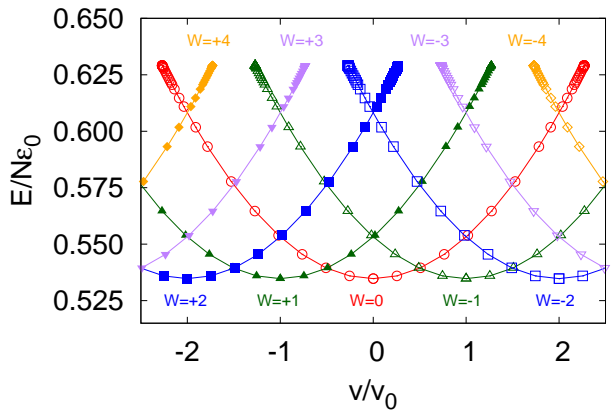}
\caption{ (Color online) Energy diagram for the stable or metastable branches for $(L, U_0, d)=(32\xi, 5\epsilon_0$, $2.5\xi$). The red, green, blue, purple, and orange lines correspond to branches with a winding number $|W|=0, 1, 2, 3$, and 4, respectively. Open (solid) symbols represent energy branches with a nonpositive (positive) winding number.}
\label{fig:energy-diagram_32_5_25}
\end{figure}%

First, we show the energy diagram, which represents the total energy in the moving frame as a function of driving velocity of moving defect. The energy diagram  yields the superfluid fraction (the nonclassical rotational inertia) \cite{Lieb2002_2}, the critical velocity, and the metastability and hysteresis of superflow states \cite{Mueller2002}.

The total energy in the moving frame is defined by
\begin{align}
E&=\int d\bm{r}\left[\frac{\hbar^2}{2m}|\nabla\Psi(\bm{r})|^2+U(\bm{r})|\Psi(\bm{r})|^2+\frac{g}{2}|\Psi(\bm{r})|^4\right].\label{eq:Results_total_energy_in_moving_frame}
\end{align}
Figure~\ref{fig:energy-diagram_32_5_25} shows the results for the stable branches. The total energy is periodic with respect to the driving velocity $v$; it stems from the periodicity of the boundary condition (\ref{eq:Model_twisted_periodic_boundary_condtion_x}).

The lowest energy state under a given $v$ is the ground state, and the other states are metastable. We confirm the metastability by calculating the excitation spectra of the Bogoliubov equation around each stationary state of the GP equation (see Sec.~\ref{subsec:excitation_spectra_2D} for details). The energy branches shown in Fig.~\ref{fig:energy-diagram_32_5_25} are almost parabolic, except in the vicinity of the termination points, which correspond to the critical velocity $v_{\rm c}$. Each branch can be specified by the winding number $W$, which we defined by (\ref{eq:Model_definition_of_winding_number}). For example, the red branch denoted by open circles ($\circ$), which continuously connects the ground state at $v=0$, has $W=0$. The green branch, denoted by open triangles ($\triangle$), has $W=-1$.

The winding number can serve as an {\it adiabatic invariant} under an adiabatic change of $v$ \cite{Mueller2002,Liu2003}. Suppose that the system is in the ground state ($W=0$), and $v$ increases adiabatically from $0$ to $v_{\rm c}$. We then expect that the system will evolve along the red branch, and the winding number will remain unchanged. Ring trap experiments \cite{Wright2013,Wright2013_full} used the ground state of the noncirculating state as the initial condition in order to see the dynamics under a change of $v$.

For the GP equation, the adiabatic condition is determined by the Bogoliubov spectrum \cite{Liu2003,Wu2003}; in the present case, the dynamics is regarded as adiabatic when there is little change in $v$ within the time interval $\hbar/\Delta$ (where $\Delta$ denotes the lowest excitation energy of the Bogoliubov spectrum). As we will show in Sec.~\ref{subsec:excitation_spectra_2D}, $\Delta$ vanishes at $v=v_{\rm c}$. When $v$ approaches $v_{\rm c}$, the adiabaticity condition is violated at a certain $v$, and a transition from $W=0$ to a circulating state $(W=-1)$ occurs, which corresponds to a phase slip \cite{Anderson1966,Langer1967}.

\begin{figure}[t]
\centering
\includegraphics[width=8.0cm,clip]{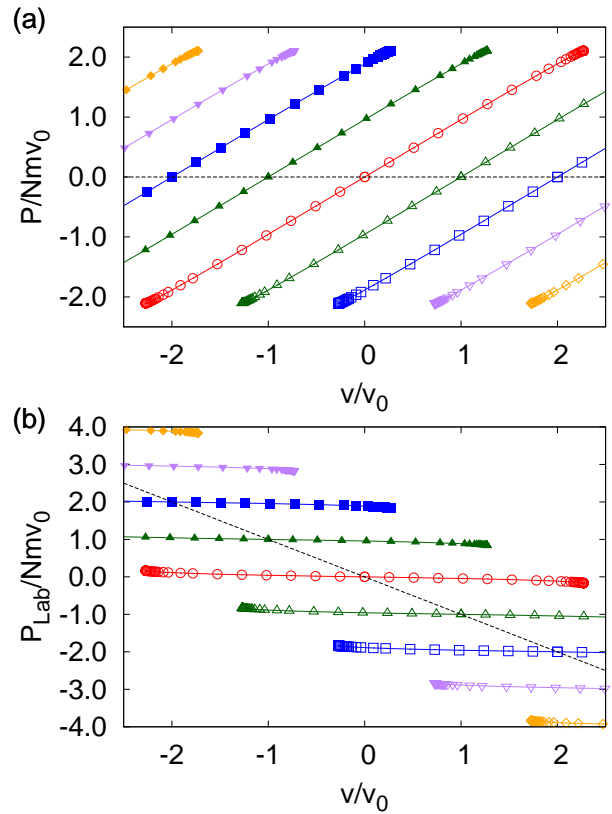}
\caption{ (Color online) Velocity dependence of the $x$ component of the total momentum per particle for the stable branches [$(L, U_0, d)=(32\xi, 5\epsilon_0, d=2.5\xi$)] in (a) the moving frame and (b) the laboratory frame. The red, green, blue, purple, and orange lines correspond to branches with winding numbers $|W|=0, 1, 2, 3$, and 4, respectively. Open (solid) symbols represent nonpositive (positive) winding numbers. The dashed black lines in (a) and (b) indicate $P=0$ and $P_{\rm Lab}=-N m v$, respectively, which represent the velocity dependence of normal fluids.}
\label{fig:momentum_velocity_32_5_25}
\end{figure}

The flow properties of each branch can be seen in the velocity dependence of the total momentum shown in Figs.~\ref{fig:momentum_velocity_32_5_25}(a) and \ref{fig:momentum_velocity_32_5_25}(b), where the total momentum of the moving frame and laboratory frame are, respectively, given by
\begin{align}
P&\equiv -\frac{i\hbar}{2}\int d\bm{r}\left[\Psi^{\ast}(\bm{r})\frac{\partial}{\partial x}\Psi(\bm{r})-\Psi(\bm{r})\frac{\partial}{\partial x}\Psi^{\ast}(\bm{r})\right],\label{eq:Results_definition_of_total_momentum}\\
P_{\rm Lab}&=P-N m v.\label{eq:Results_definition_of_total_momentum_lab_frame}
\end{align}
The $y$-component of the total momentum is zero by symmetry, and we thus consider only the $x$-component. For the red branch, the total momentum of the moving frame has a linear dependence for small $v$; see Fig.~\ref{fig:momentum_velocity_32_5_25}(a). The superfluid fraction can be calculated through the following relations\cite{Baym1969,Lieb2002_2}:
\begin{eqnarray}
\frac{\rho_{\rm s}}{\rho}\equiv \frac{1}{N m}\left.\frac{\partial P(v)}{\partial v}\right|_{v\to 0}=\frac{1}{Nm}\left.\frac{\partial^2E(v)}{\partial v^2}\right|_{v \to 0},\label{eq:Results_definition_of_superfluid_fraction}
\end{eqnarray}
where we used the relation $P(v)=\partial E(v)/\partial v$. The calculated value of the superfluid fraction is $\rho_{\rm s}/\rho=0.9598298(5)$ for $(L, U_0, d)=(32\xi, 5\epsilon_0, 2.5\xi)$ \cite{note_fraction}. If we consider a uniform system, the superfluid fraction becomes unity because the total momentum is given by $P(v)=N m v$. The deviation from unity for our system is due to the presence of the external potential. From Eq.~(\ref{eq:Results_definition_of_superfluid_fraction}) and Ref.~\cite{Kramer2003}, we can show that the effective mass $m^{\ast}(v)$ at $v=0$ is related to $m/m^{\ast}(v=0)=\rho_{\rm s}/\rho$. In the presence of the external potential, usually $m^{\ast}(v=0)/m>1$ holds and hence $\rho_{\rm s}/\rho<1$.

\begin{figure}[t]
\centering
\includegraphics[width=8.0cm,clip]{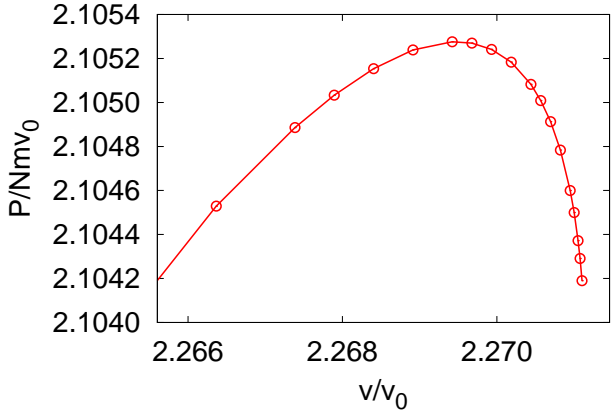}
\caption{(Color online) Magnified view of the $W=0$ branch in Fig.~\ref{fig:momentum_velocity_32_5_25}(a) near the critical velocity.}
\label{fig:momentum_velocity_zoom_32_5_25}
\end{figure}

Figure~\ref{fig:momentum_velocity_zoom_32_5_25} shows a blow-up of the region near the critical velocity for the $W=0$ branch. We note that the effective mass [$(m/m^{\ast})=(1/Nm)\partial P(v)/\partial v<0$] becomes negative near the critical velocity; this implies that the mass flow of the condensate in the moving frame decreases while the velocity of the moving defect becomes larger. Negative effective-mass states have been found in the GP equation for a BEC in an optical lattice near the critical velocity (see Fig.~7 in Ref.~\cite{Danshita2007}). However, negative effective-mass states in an optical lattice are subject to dynamical instability (DI) \cite{Machholm2003,Wu2003,Menotti2003_2,Taylor2003,Morsch2006,Danshita2007}, while they are metastable in our case. This difference comes from that the DI in the optical lattice systems is due to the formation of the long-period structures such as period-doubling solutions \cite{Machholm2004} or bright gap solitons \cite{Konotop2002,Eiermann2004}. These structures are prohibited in a torus and thus the negative-effective mass states maintain metastability.

We note that the qualitatively same behavior for negative effective-mass states is found for other values of the parameters, as follows: $L/\xi=24,32,48,64$, $U/\epsilon_0=1,10,20$, and $d/\xi=1, 2,2.5,5$. We thus believe that the results in this section are generic for a superfluid in a torus near the critical velocity in the presence of a moving defect. 

%%%%%%%%%%%%%%%%%%%%%%%%%%%%%%%%%%%%%%%%%%%%%%%%%%%%%%%%%%%%%%
\subsection{Density and phase profile}\label{subsec:density_and_phase_profile}
%%%%%%%%%%%%%%%%%%%%%%%%%%%%%%%%%%%%%%%%%%%%%%%%%%%%%%%%%%%%%%

We next consider the spatial profiles of the density and the phase of the condensate wave function for a strong potential $U_0=10\epsilon_0$ and a weak potential $U_0=\epsilon_0$.

\begin{figure}[t]
\centering
\includegraphics[width=7.0cm,clip]{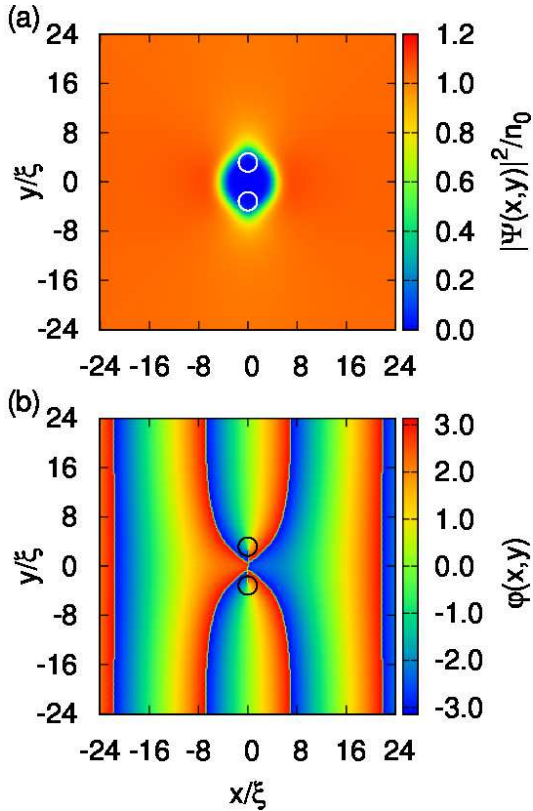}
\caption{(Color online) (a) Density profile, and (b) phase profile, for $(L, U_0,d)=(48\xi, 10\epsilon_0, 2.5\xi)$ and $v=0.4303400v_{\rm s}\simeq 0.9999907 v_{\rm c}$. White and black circles represent the position of the GVP.}
\label{fig:density_and_phase_profile_48_10_25_0.43034}
\end{figure}

\begin{figure}[t]
\includegraphics[width=7.0cm,clip]{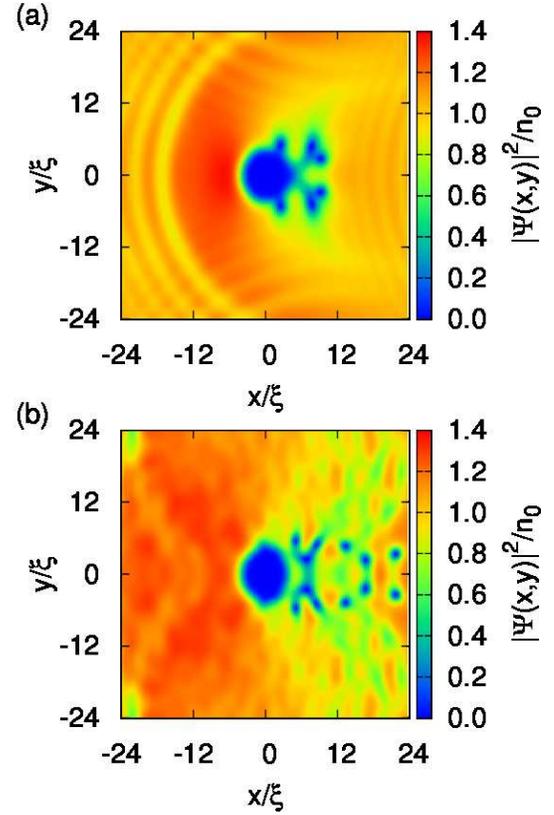}
\caption{(Color online) Snapshot of the density for $(L, U_0, d)=(48\xi, 10\epsilon_0, 2.5\xi)$ and $v=0.8v_{\rm s}\simeq 1.8590v_{\rm c}$, at (a) $t=20\tau$ and (b) $t=50\tau$. The initial condition is the stationary solution for $v=0.4v_{\rm s}\simeq 0.92950v_{\rm c}$, which contains the GVP.}
\label{fig:density_48_10_25_04_08_t20_t50}
\end{figure}

Figures ~\ref{fig:density_and_phase_profile_48_10_25_0.43034}(a) and \ref{fig:density_and_phase_profile_48_10_25_0.43034}(b) show 
the spatial profiles of the density and phase, respectively, of the condensate wave function for $(L, U_0, d)=(48\xi, 10\epsilon_0, 2.5\xi)$ and $v=0.4303400v_{\rm s}\simeq 0.9999907 v_{\rm c}$. We find a vortex pair in the low-density region; this is called a ghost vortex pair (GVP) \cite{Tsubota2002,Kasamatsu2003,Fujimoto2011}. This is a (meta)stable stationary solution, because no DI occurs in the solution (see Sec.~\ref{subsec:excitation_spectra_2D}). The GVP is regarded as being pinned to the defect potential, and thus it could be depinned above the critical velocity. In fact, we calculated the real-time dynamics above the critical velocity and found that vortex nucleation occurred as shown in Figs.~\ref{fig:density_48_10_25_04_08_t20_t50}(a) and \ref{fig:density_48_10_25_04_08_t20_t50}(b). We generated the real-time dynamics by the Crank-Nicholson scheme.

Ghost vortices were first reported in Refs. \cite{Tsubota2002,Kasamatsu2003}, where vortex invasions of rotating condensates were numerically studied. A GVP was found in a numerical study of condensates in the presence of an oscillating defect \cite{Fujimoto2011}. Our result yields an example of a GVP accompanying a defect moving with a constant subcritical velocity.

\begin{figure}[t]
\includegraphics[width=7.0cm,clip]{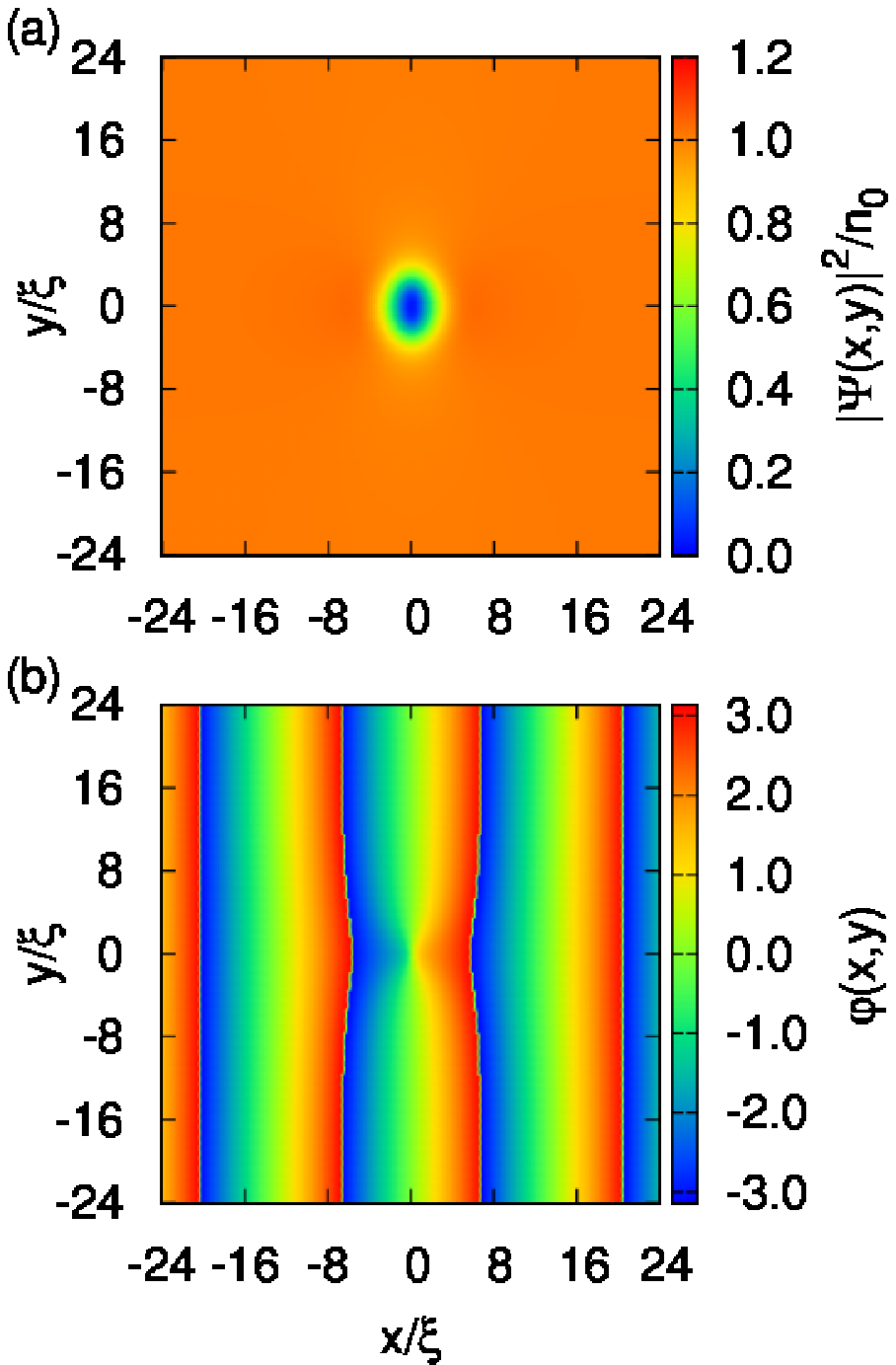}
\caption{(Color online) (a) Density profile, and (b) phase profile, for $(L, U_0, d)=(48\xi, \epsilon_0, 2.5\xi)$ and $v=0.4608505v_{\rm s}\simeq 0.9999984 v_{\rm c}$.}
\label{fig:density_and_phase_profile_48_1_25_0.4608505}
\end{figure}

\begin{figure}[t]
\includegraphics[width=7.0cm,clip]{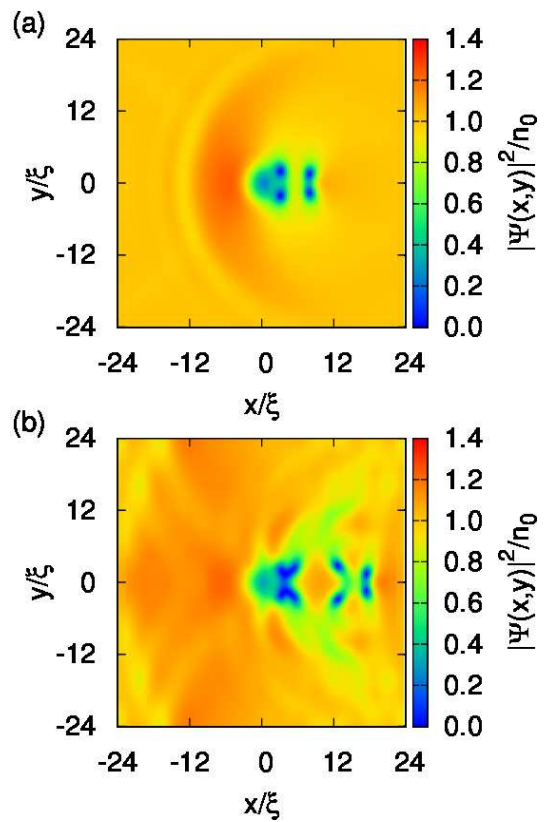}
\caption{(Color online) Snapshot of the density for $(L, U_0, d)=(48\xi, \epsilon_0, 2.5\xi)$ and $v=0.8v_{\rm s}\simeq 1.73590v_{\rm c}$, at (a) $t=20\tau$, and (b) $t=50\tau$. The initial condition is the stationary solution for $v=0.4v_{\rm s}\simeq 0.867959v_{\rm c}$.}
\label{fig:density_48_1_25_04_08_t20_t50}
\end{figure}

A GVP does not appear in the presence of a weak potential; this is shown in Figs.~\ref{fig:density_and_phase_profile_48_1_25_0.4608505}(a) and \ref{fig:density_and_phase_profile_48_1_25_0.4608505}(b), where we present the density and phase profiles for $(L, U_0, d)=(48\xi, \epsilon_0, 2.5\xi$) and $v=0.4608505v_{\rm s}\simeq 0.9999984v_{\rm c}$. Typically, investigations are for a velocity in the range of $10^{-6}\lesssim |(v_{\rm c}-v)/v_{\rm c}|\le 1$. The dynamics above the critical velocity in the weak-potential case are shown in Figs.~\ref{fig:density_48_1_25_04_08_t20_t50}(a) and \ref{fig:density_48_1_25_04_08_t20_t50}(b). These figures clearly show that vortices nucleate even when the initial state contains no GVPs.  

%%%%%%%%%%%%%%%%%%%%%%%%%%%%%%%%%%%%%%%%%%%%%%%%%%%%%%%%%%%%%%
\subsection{Excitations}\label{subsec:excitation_spectra_2D}
%%%%%%%%%%%%%%%%%%%%%%%%%%%%%%%%%%%%%%%%%%%%%%%%%%%%%%%%%%%%%%

\begin{figure}[t]
\centering
\includegraphics[width=8.0cm,clip]{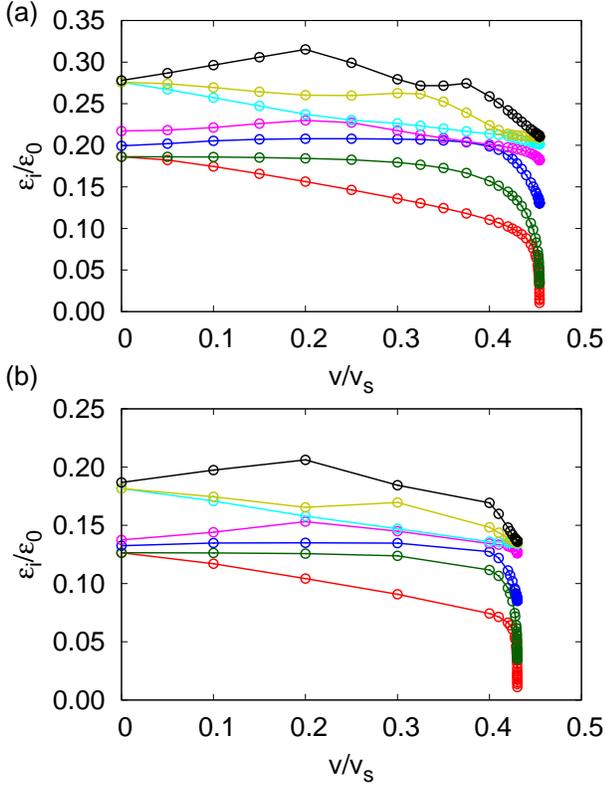}
\caption{(Color online) Velocity dependence of the excitation energy for (a) $(L, U_0, d)=(32\xi, 10\epsilon_0, 2.5\xi)$; and (b) $(L, U_0,d)=(48\xi, 10\epsilon_0, 2.5\xi)$. Up to the seventh excitation energy is shown.}
\label{fig:excitation_all_32_and_48}
\end{figure}% 
\begin{figure}[t]
\centering
\includegraphics[width=8.0cm,clip]{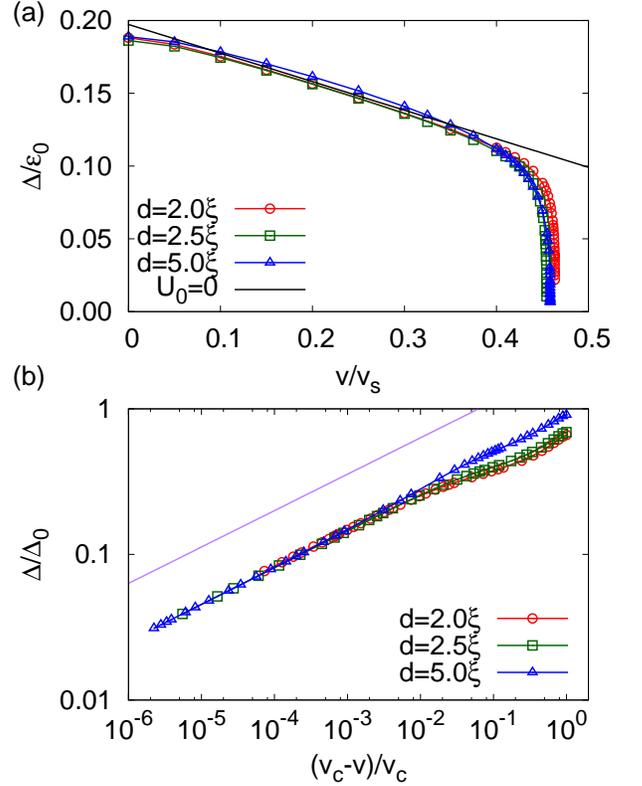}
\caption{(Color online) (a)Velocity dependence of the energy gap for $L=32\xi$ and $U_0=10\epsilon_0$. The solid black line shows the energy gap for $U_0=0$. (b) Fitting results for (a). The solid purple line represents $[(v_{\rm c}-v)/v_{\rm c}]^{1/4}$.}
\label{fig:gap_scaling_32_10_d}
\end{figure}%
\begin{figure}[t]
\centering
\includegraphics[width=8.0cm,clip]{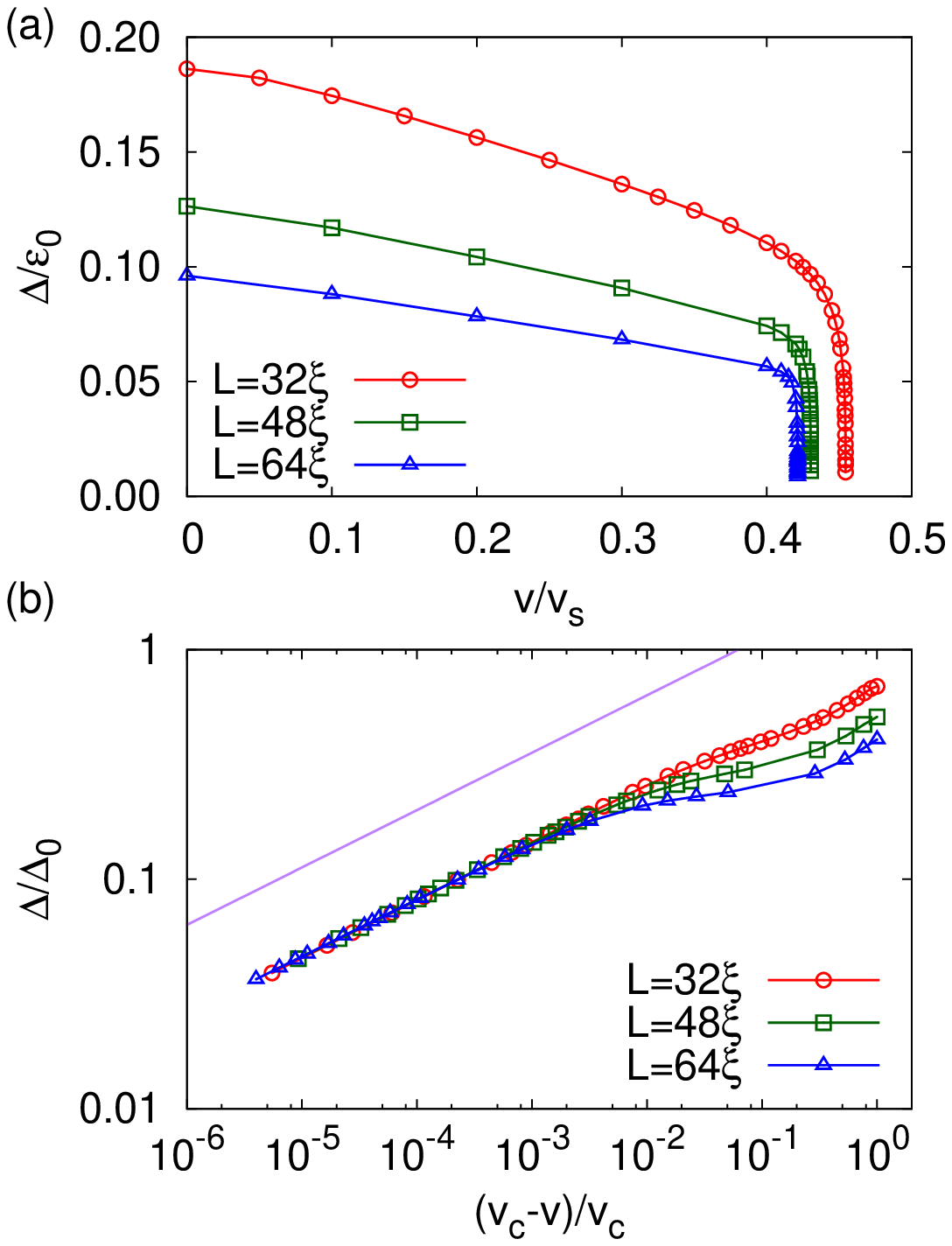}
\caption{(Color online) (a) Velocity dependence of the energy gap for $U_0=10\epsilon_0$ and $d=2.5\xi$. (b) Fitting results for (a). The solid purple line represents $[(v_{\rm c}-v)/v_{\rm c}]^{1/4}$.}
\label{fig:gap_scaling_32_48_64_10_25}
\end{figure}%

We next present the results for excitations. Figure~\ref{fig:excitation_all_32_and_48} shows  energy spectra of the Bogoliubov equation as a function of $v$. In systems of finite size, the Bogoliubov spectra are discretized. The excitation energy is always positive, and thus the solutions shown in the figure are stable or metastable.

We now focus on the first excited energy, which we call an energy gap (denoted by $\Delta $). Figures~\ref{fig:gap_scaling_32_10_d}(a) and \ref{fig:gap_scaling_32_48_64_10_25}(a) show the energy gap as a function of the velocity. 

We first notice  a linear decrease in the region in which the velocity is small. This reflects the energy gap in uniform systems, and it is given by $\Delta_{\rm uni}/\epsilon_0=2\pi/(L/\xi)\left[-v/v_{\rm s}+\sqrt{\pi^2/(L/\xi)^2+1}\right]$. In fact, the solid black line in Fig.~\ref{fig:gap_scaling_32_10_d}(a), which represents $\Delta_{\rm uni}$, almost overlaps the numerical data for $U_0\not=0$ when the velocity is small, except near $v=0$. The deviation between $\Delta_{\rm uni}$ and the numerical data near $v=0$ is due to the splitting of levels, since the first excited state in a uniform system is fourfold degenerate. 

We also notice a sharp decrease in the energy gap near the critical velocity. To characterize this behavior, we used the function $\Delta=\Delta_0[(v_{\rm c}-v)/v_{\rm c}]^{c}$, where $\Delta_0, v_{\rm c}$, and $c$ are parameters to fit four sets of data points near the critical velocity. The detailed data are shown in Table 1 of Ref.~\cite{Kunimi2014}. Figures \ref{fig:gap_scaling_32_10_d}(b) and \ref{fig:gap_scaling_32_48_64_10_25}(b) show the  results. We determined that the scaling for the energy gap $\Delta=\Delta_0[(v_{\rm c}-v)/v_{\rm c}]^{1/4}$ near the critical velocity.
 
%%%%%%%%%%%%%%%%%%%%%%%%%%%%%%%%%%%%%%%%%%%%%%%%%%%%%%%%%%%%%%
\subsection{Fluctuations}\label{subsec:fluctuation_2D}
%%%%%%%%%%%%%%%%%%%%%%%%%%%%%%%%%%%%%%%%%%%%%%%%%%%%%%%%%%%%%%

So far we have shown only the excitation spectra. We will show the wave functions for excited states in this subsection.

Using the wave functions of the excited states, we can obtain the properties of the fluctuations. Substituting Eq.~(\ref{eq:Model_linearizing_GP_equation}) into $n(\bm{r}, t)=|\Psi(\bm{r}, t)|^2$ and $\Psi(\bm{r}, t)/|\Psi(\bm{r}, t)|$, and neglecting the higher-order terms of $u_i(\bm{r})$ and $v_i(\bm{r})$, we obtain
\begin{align}
n(\bm{r}, t)&=|\Psi(\bm{r})|^2+2{\rm Re}\left[\delta n_i(\bm{r})e^{-i\epsilon_it/\hbar}\right],\label{eq:results_density_fluctuation}\\
\frac{\Psi(\bm{r}, t)}{|\Psi(\bm{r}, t)|}&=e^{-i\mu t/\hbar}e^{i\varphi(\bm{r})}\nonumber \\
&\quad \times \left\{1+\frac{i}{|\Psi(\bm{r})|^2}{\rm Im}\left[\delta P_i(\bm{r})e^{-i\epsilon_it/\hbar}\right]\right\},\label{eq:results_phase_fluctuation}
\end{align}
where the local density fluctuation $\delta n_i(\bm{r})$ and the local phase fluctuation $\delta P_i(\bm{r})$ for mode $i$ are defined by \cite{Wen-Chin_Wu1996,Fetter1998_2}
\begin{eqnarray}
\delta n_i(\bm{r})&=&\Psi^{\ast}(\bm{r})u_i(\bm{r})-\Psi(\bm{r})v_i(\bm{r}),\label{eq:results_local_density_fluctuation}\\
\delta P_i(\bm{r})&=&\Psi^{\ast}(\bm{r})u_i(\bm{r})+\Psi(\bm{r})v_i(\bm{r}).\label{eq:results_local_phase_fluctuation}
\end{eqnarray}

\begin{figure*}[t]
\centering
\includegraphics[width=15.0cm,clip]{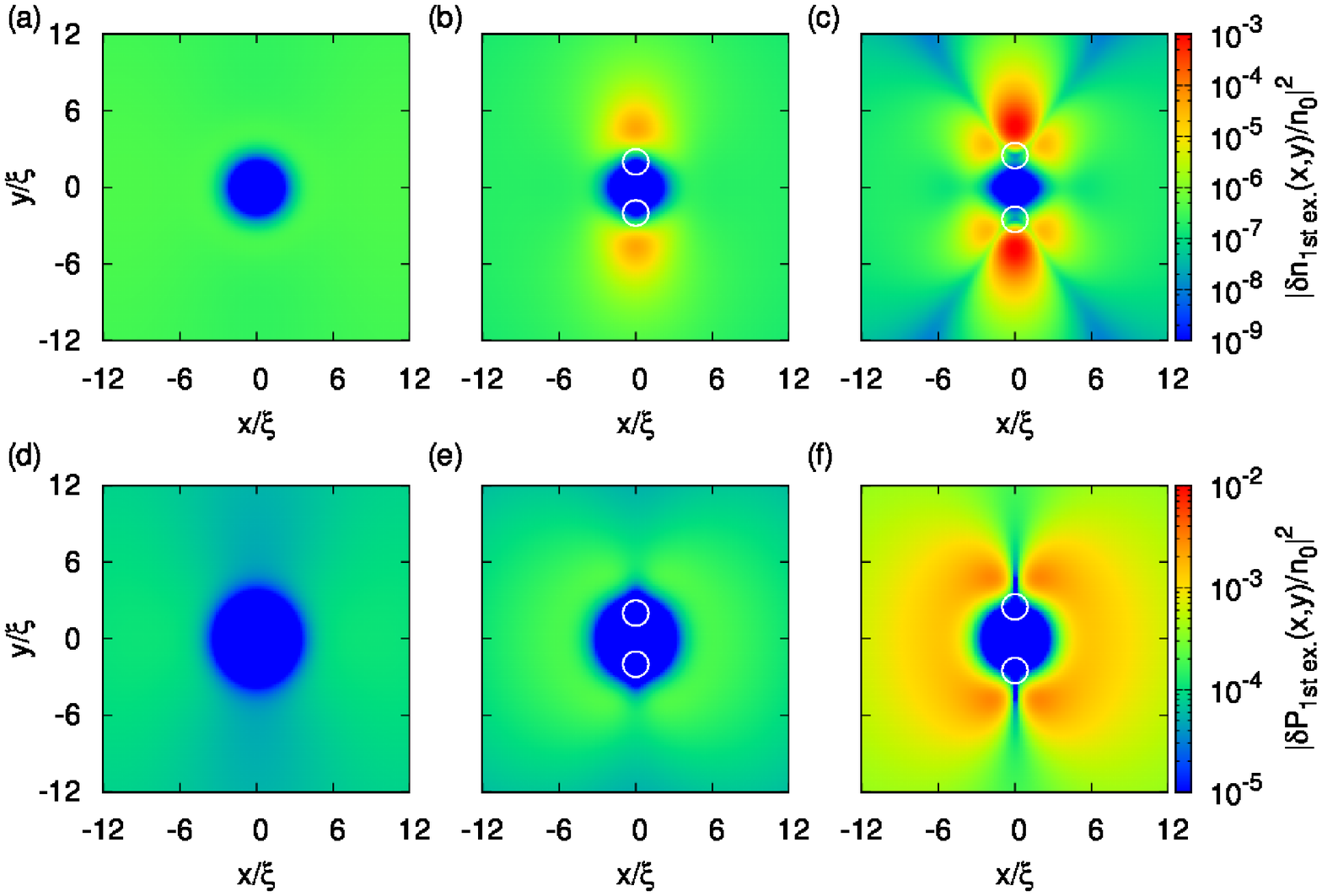}
\caption{(Color online) (a)$-$(c) Spatial profiles of the density fluctuation for $(L, U_0,d)=(48\xi, 5\epsilon_0, 2.5\xi)$ by the first excited state for $v=0.1v_{\rm s}, 0.42v_{\rm s},$ and $v=0.42655v_{\rm s}$, respectively. (d)$-$(f) Spatial profiles of the phase fluctuation for $(L, U_0,d)=(48\xi, 5\epsilon_0, 2.5\xi)$, for the first excited state for $v=0.1v_{\rm s}$, $v=0.42v_{\rm s},$ and $v=0.42655v_{\rm s}$, respectively. The white circles represent the position of the GVP. Here, we set $1/\sqrt{n_0\xi^2}=0.1$.}
\label{fig:fluctuation_density_and_phase_48_5_25}
\end{figure*}

\begin{figure*}[t]
\centering
\includegraphics[width=15.0cm,clip]{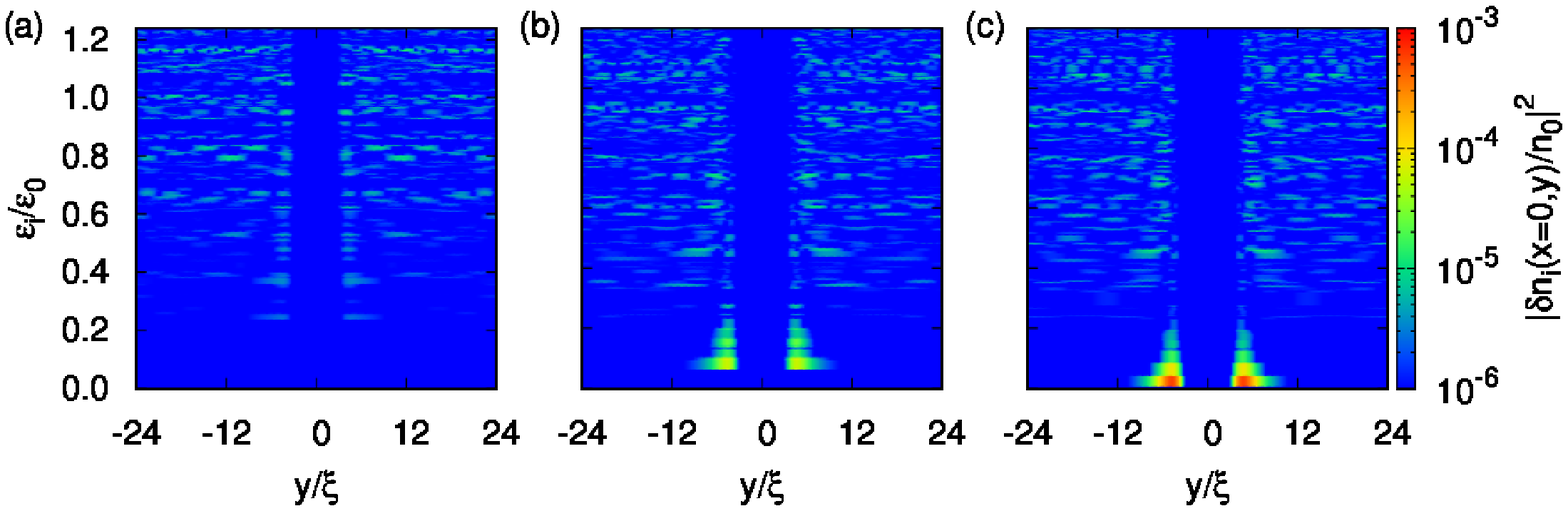}
\caption{(Color online) Energy and $y$ dependence of the density fluctuations for $(L, U_0, d)=(48\xi, 5\epsilon_0, 2.5\xi)$, (a) $v=0.1v_{\rm s}$, (b) $v=0.42v_{\rm s},$ and (c) $v=0.42655v_{\rm s}$, at $x=0$. Here, we set $1/\sqrt{n_0\xi^2}=0.1$.}
\label{fig:fluctuation_energy_48_5_25}
\end{figure*}

In Fig.~\ref{fig:fluctuation_density_and_phase_48_5_25}, we show the spatial profiles of the density and phase fluctuations for the first excited state. We can see that both the density and phase fluctuations are enhanced when the velocity of the moving defect becomes larger. The enhancement for the density fluctuation is a few orders of magnitude greater than that for the phase fluctuation. We plot the energy dependence of the density fluctuations in Fig.~\ref{fig:fluctuation_energy_48_5_25}, where the spectral intensity shifts to lower energy and is enhanced when the velocity approaches the critical value. Similar behavior was observed in one-dimensional systems, and was related to soliton nucleation \cite{Kato2010,Watabe2013}.

We briefly remark on the effect of the existence or absence of a GVP on the fluctuations. For the weak potential case, there is no GVP near the critical velocity, as described in Sec.~\ref{subsec:density_and_phase_profile}. In this case, the spatial patterns of the density and phase fluctuations are slightly different (data not depicted here). However, enhancement of the density fluctuation also occurs in the absence of a GVP. 

%%%%%%%%%%%%%%%%%%%%%%%%%%%%%%%%%%%%%%%%%%%%%%%%%%%%%%%%%%%%%%
\subsection{Quantum depletion}\label{subsec:Quantum depletion}
%%%%%%%%%%%%%%%%%%%%%%%%%%%%%%%%%%%%%%%%%%%%%%%%%%%%%%%%%%%%%%

We will present the results for quantum depletion (QD), which represents the number of atoms not in the condensate. The expression for the QD is
\begin{eqnarray}
\frac{N_{\rm dep}}{N}&=&\frac{1}{N}\sum_{i}\int d\bm{r}|v_i(\bm{r})|^2.\label{eq:resulsts_quantum_depletion}
\end{eqnarray}
The condition for the GP and Bogoliubov approximations to be valid is $N_{\rm dep}/N\ll 1$. Therefore, we can check the self-consistency of these approximations by examining the QD.

In previous works, QD in nonuniform systems was calculated in the following cases: These studies used perturbative approaches \cite{Huang1992,Kobayashi2002,Muller2012} and many-body calculations for the ground state \cite{Astrakharchik2011,Astrakharchik2013}. Our results presented below are the first examples of the QD near the critical velocity in the presence of a moving defect.

\begin{figure}[t]
\centering
\includegraphics[width=8.0cm,clip]{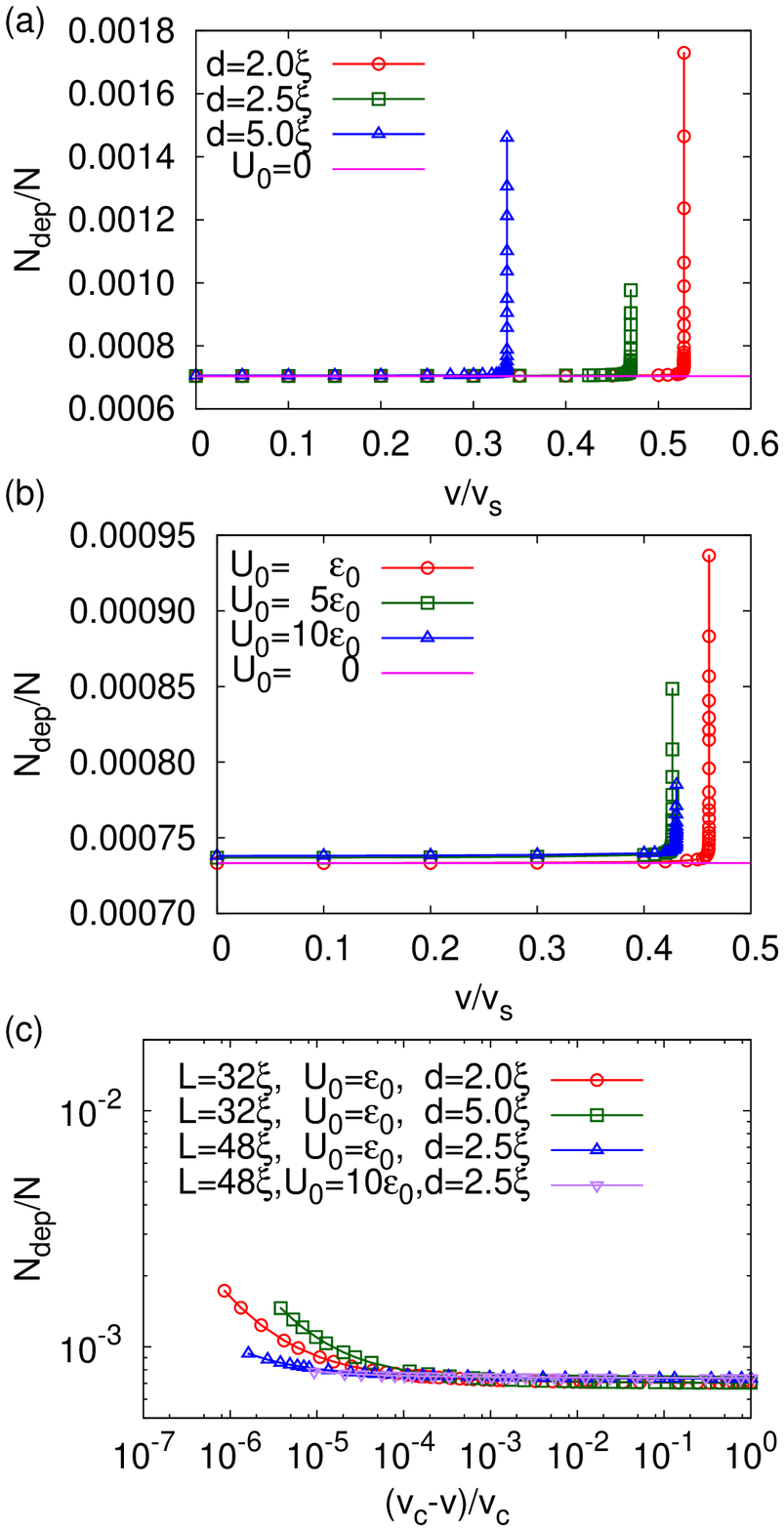}
\caption{(Color online) Velocity dependence of the QD for (a)$L=32\xi$ and $U_0=\epsilon_0$, and (b)$L=48\xi$ and $d=2.5\xi$. The solid purple line represents the QD in the absence of the defect potential. (c) Normalized velocity dependence of the QD. In all cases, we set $1/\sqrt{n_0\xi^2}=0.1$.}
\label{fig:depletion_32_and_48}
\end{figure}%

We show the velocity dependence of the QD in Fig.~\ref{fig:depletion_32_and_48}  \cite{Note_depletion}. If the system is uniform ($U=0$) and does not exhibit spontaneous translational symmetry breaking, the QD does not depend on the velocity because the velocity dependence of $v_i(\bm{r})$ is present only in the plane wave component ($e^{-i m\bm{v}\cdot\bm{r}/\hbar}$). Our results show that the QD depends on the velocity, due to the presence of the defect potential. When the velocity is small, the QD is almost the same as that for uniform systems. It is consistent with that for the energy gap when the velocity is small. Near the critical velocity, we find that the QD increases steeply and attribute it to enhancement of the low-energy density of states and density fluctuations. Within the range of $v$ used in our calculations, we do not find power-law scaling of the QD, in contrast to that found for the energy gap; see Fig.~\ref{fig:depletion_32_and_48} (c), and note that the curves are not straight lines in the log-log plot near the critical velocity. There may be a narrow scaling region for the QD in this system.

\begin{figure}[t]
\centering
\includegraphics[width=8.0cm,clip]{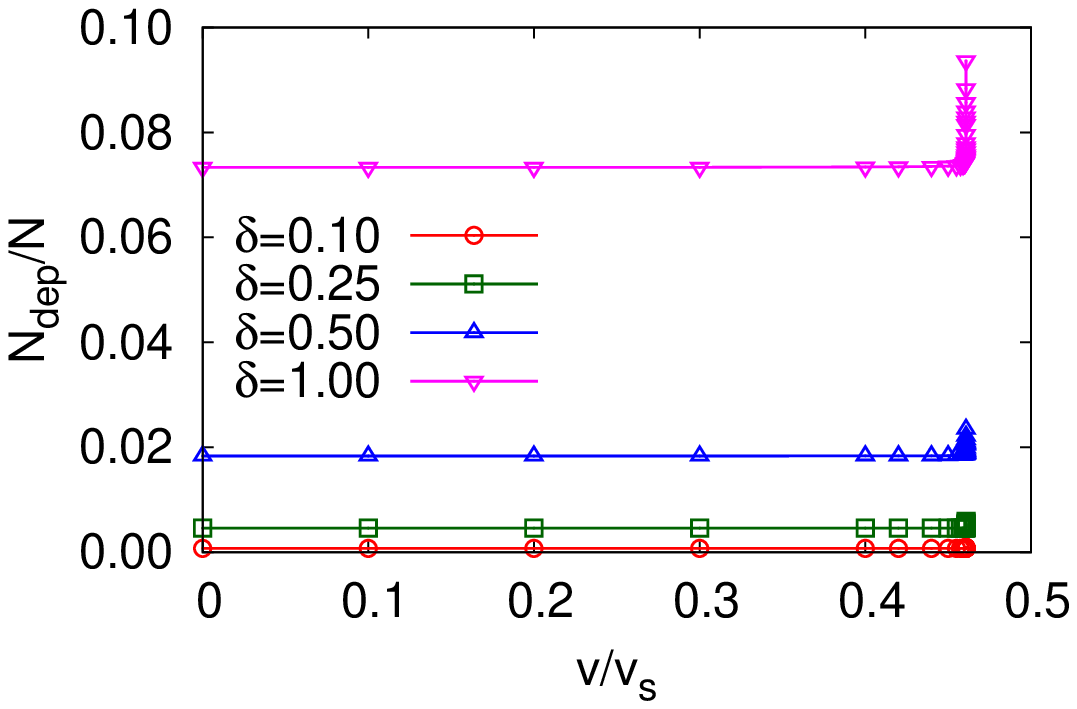}
\caption{(Color online) Velocity dependence of the quantum depletion for $(L, U_0, d)=(48\xi, \epsilon_0, 2.5\xi)$, for four different values of $\delta$.}
\label{fig:dep_different_delta_48_1_25}
\end{figure}

Although the QD increases near the critical velocity, the value of $N_{\rm dep}/N$ is still much smaller than unity for $\delta\equiv 1/\sqrt{n_0\xi^2}=0.1$ and $(v_{\rm c}-v)/v_{\rm c}\gtrsim 10^{-6}$; see Fig.~\ref{fig:depletion_32_and_48} (c). Here, $\delta$ is the ratio between the healing length and the mean particle distance. A small $\delta$ corresponds to a weakly interacting case. This shows that the GP and Bogoliubov approximations are valid even near the critical velocity, for sufficiently small $\delta$. We can easily obtain the QD for other values of $\delta$, because the $\delta$ dependence of $v_i(\bm{r})$ is given by $v_i(\bm{r} ; \delta)=\delta\times v_i(\bm{r} ; \delta=1)$. Figure \ref{fig:dep_different_delta_48_1_25} shows the $\delta$ dependence of the QD. These results show that the QD is much smaller than unity near the critical velocity for $\delta\le 0.5$, and for $\delta>1$, the Bogoliubov approximation breaks down near the critical velocity. 

%%%%%%%%%%%%%%%%%%%%%%%%%%%%%%%%%%%%%%%%%%%%%%%%%%%%%%%%%%%%%%
\section{Results for Unstable Branches}\label{sec:Results_unstable}
%%%%%%%%%%%%%%%%%%%%%%%%%%%%%%%%%%%%%%%%%%%%%%%%%%%%%%%%%%%%%%

In this section, we will present the results for unstable branches, which we calculated by the pseudo-arclength continuation method \cite{Keller1986}. The details are summarized in Appendix~\ref{app:PACM}.

Figure \ref{fig:energy_diagram_sw_32_5_25} shows the energy diagram containing the stable and the unstable branches for $(L, U_0, d)=(32\xi, 5\epsilon_0,2.5\xi)$. Our results do not exhibit the conventional swallowtail structure but a multiple structure of the unstable branches, in contrast to one-dimensional lattices \cite{Mueller2002,Machholm2003,Wu2003,Menotti2003_2,Taylor2003,Machholm2004,Morsch2006,Danshita2007} and ring systems \cite{Kanamoto2009,Fialko2012,Baharian2013}. We call it the multiple-swallowtail structure. The branch (a) in the inset of Fig.~\ref{fig:energy_diagram_sw_32_5_25} continuously connects the stable branch with $W=0$ at the critical velocity and contains one vortex pair as shown in Fig.~\ref{fig:density_and_phase_swallow_tail_32_5_25} (a). The upper (lower) vortex has negative (positive) vorticity. This solution is similar to the unstable solution reported in Ref.~\cite{Huepe2000}. At the left termination point, the branch (a) folds back and connects with the branch (b) in the energy diagram. We have two pairs of vortices in the branch (b). At $v=v_0/2$ on the top of the unstable branch (d), the self-induced phase slip \cite{Kanamoto2008,Kanamoto2009} occurs due to formation of the dark soliton [see Figs.~\ref{fig:density_and_phase_swallow_tail_32_5_25}(d) and \ref{fig:density_and_phase_swallow_tail_32_5_25}(h)]. Consequently, the winding number changes from $W=0$ to $W=-1$ \cite{Kanamoto2008, Kanamoto2009,Fialko2012}. After the self-induced phase slip, finally, the unstable branch (g) connects with the stable branch for $W=-1$.

In Fig.~\ref{fig:energy_diagram_sw_48_5_25}, we show the energy diagram for a larger system with $(L, U_0,d )=(48\xi, 5\epsilon_0, 2.5\xi)$. There are more unstable branches with more pairs of vortices compared to the case of $(L, U_0, d)=(32\xi, 5\epsilon_0,2.5\xi)$. We consider that the number of the branches of the multiple swallowtail structure is sensitive to the width of the superflow path (spatial extension in the $y$-direction). We expect that the number of the unstable branches reduces to unity in systems with a narrow path of superflow comparable to the healing length and the multiple swallowtail becomes conventional swallowtail. 

We show the excitation spectra for an unstable branch (a) and the stable branch near the critical velocity in Fig.~\ref{fig:excitation_sw}. As expected, the DI occurs in the unstable branch (a). We confirm that the DI occurs in other unstable branches (b) to (g) in Fig.~\ref{fig:energy_diagram_sw_32_5_25} (data not depicted here).

\begin{figure}[t]
\centering
\includegraphics[width=8.0cm,clip]{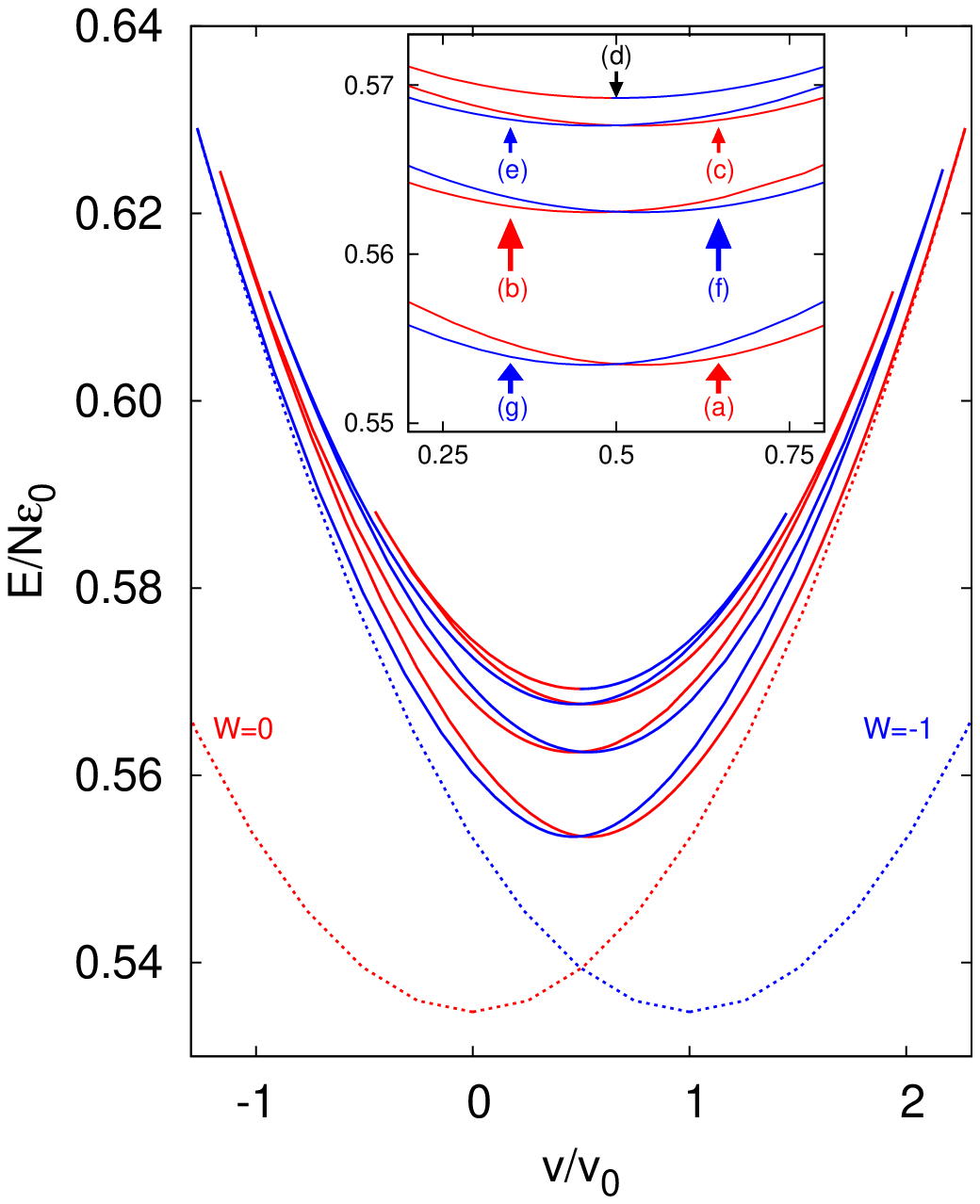}
\caption{(Color online) Energy diagram for $(L, U_0, d)=(32\xi, 5\epsilon_0, 2.5\xi)$. The red dotted (solid) line shows the stable (unstable) solution for $W=0$. The blue dotted (solid) line shows the stable (unstable) solution for $W=-1$. The inset shows magnified view around $v=v_0/2$.}
\label{fig:energy_diagram_sw_32_5_25}
\end{figure}% 

\begin{figure*}[t]
\includegraphics[width=17.5cm,clip]{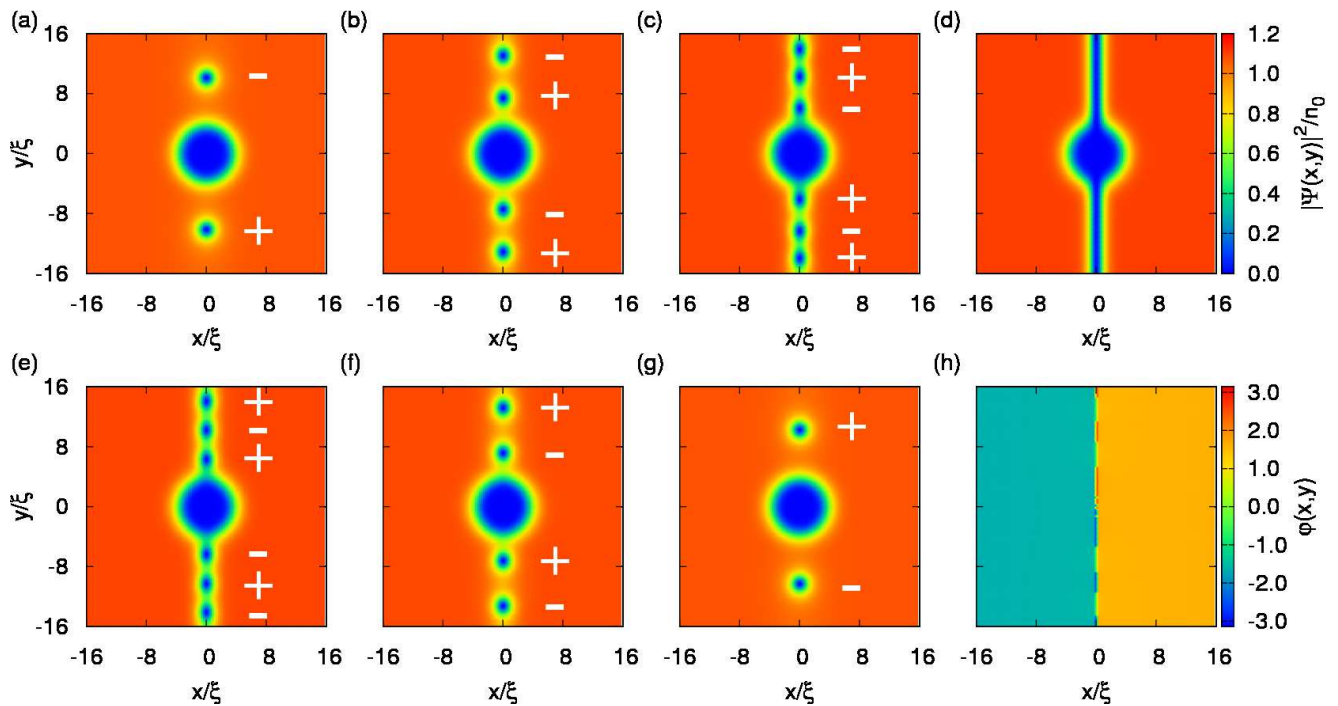}
\caption{(Color online) Density profiles for $(L, U_0, d)=(32\xi, 5\epsilon_0, 2.5\xi)$ near $v=v_0/2$. The labels (a) to (g) correspond to the branches shown in the inset of Fig.~\ref{fig:energy_diagram_sw_32_5_25}. The sign written near each vortex denotes vorticity. (h): Phase profile for (d). We can see $\pi$-phase jump at $x=0$ in (h).}
\label{fig:density_and_phase_swallow_tail_32_5_25}
\end{figure*}%

\begin{figure}[t]
\centering
\includegraphics[width=8.0cm,clip]{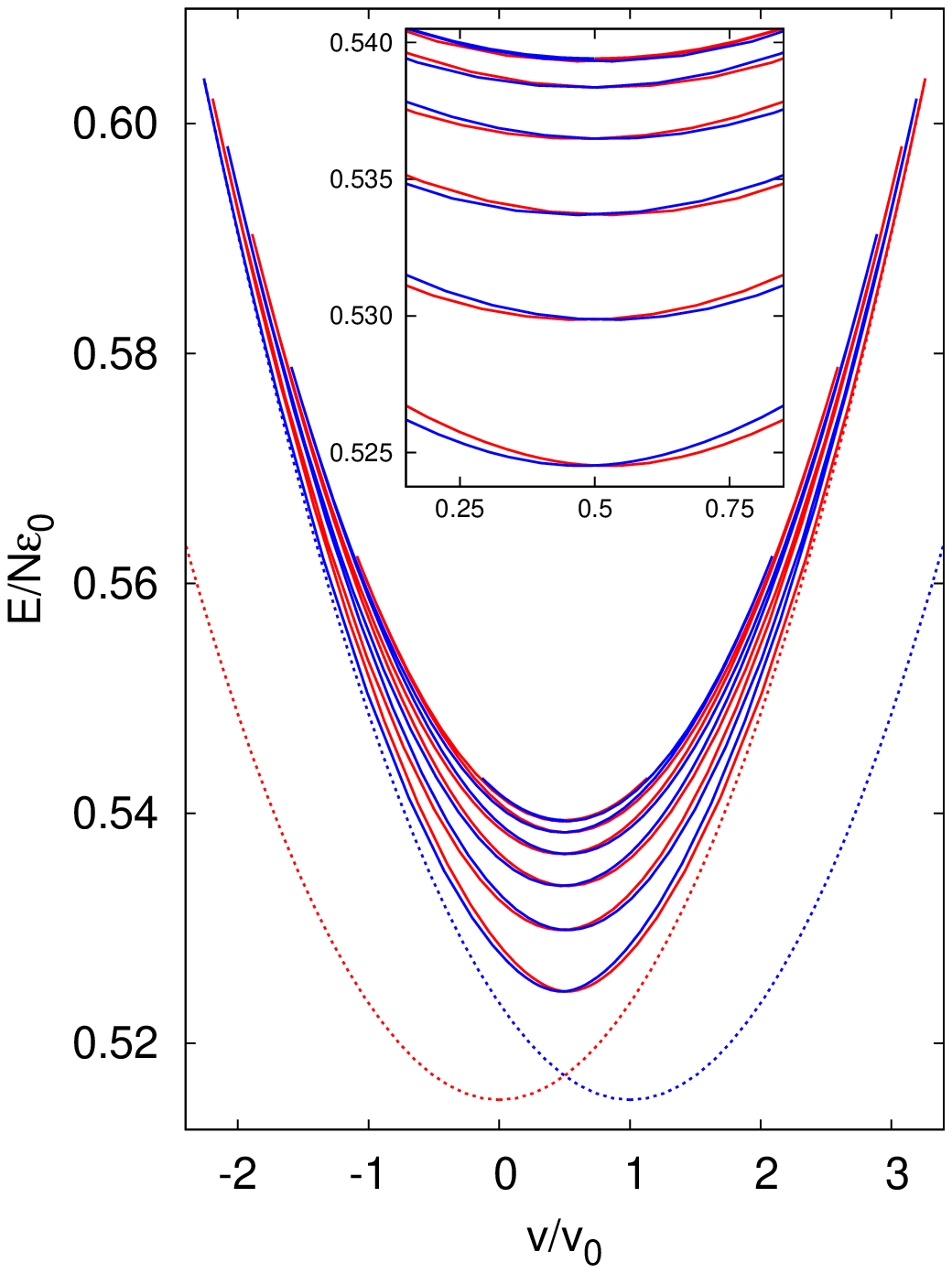}
\caption{(Color online) Energy diagram for $(L, U_0,d)=(48\xi, 5\epsilon_0, 2.5\xi)$. The red dotted (solid) line shows the stable (unstable) solution for $W=0$. The blue dotted (solid) line shows the stable (unstable) solution for $W=-1$. The inset shows magnified view around $v=v_0/2$.}
\label{fig:energy_diagram_sw_48_5_25}
\end{figure}% 

\begin{figure}[t]
\centering
\includegraphics[width=8.0cm,clip]{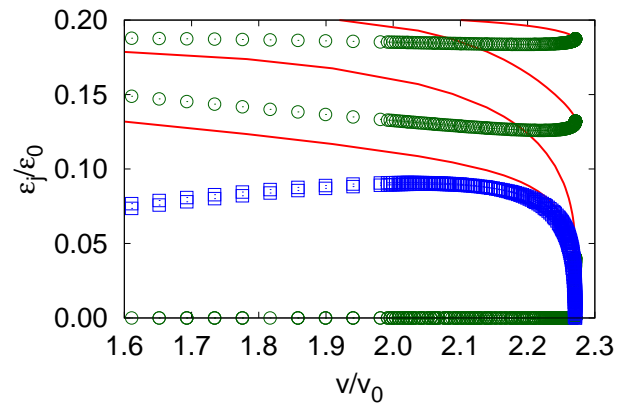}
\caption{(Color online) Velocity dependence of the excitation energy near the critical velocity for $(L, U_0,d)=(32\xi, 5\epsilon_0, 2.5\xi)$. The solid red curve represents the excitation energy of the stable branch. The green circle and the blue square represent the real and imaginary part of the excitation energy for branch (a) in Fig.~\ref{fig:energy_diagram_sw_32_5_25}, respectively.}
\label{fig:excitation_sw}
\end{figure}% 

\begin{figure}[t]
\centering
\includegraphics[width=7.0cm,clip]{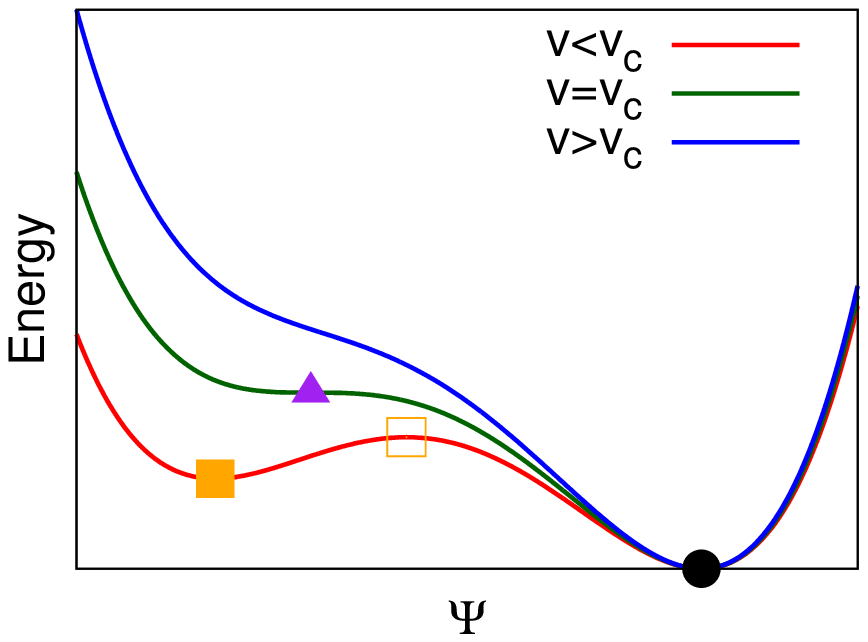}
\caption{(Color online) Schematic picture of the energy landscape in the vicinity of the critical velocity. Each local minimum and maximum corresponds to a stationary solution of the GP equation. Filled (open) symbols represent local minima (maxima).}
\label{fig:schematic_picture}
\end{figure}%

%%%%%%%%%%%%%%%%%%%%%%%%%%%%%%%%%%%%%%%%%%%%%%%%%%%%%%%%%%%%%%
\section{Discussion}\label{sec:discussion}
%%%%%%%%%%%%%%%%%%%%%%%%%%%%%%%%%%%%%%%%%%%%%%%%%%%%%%%%%%%%%%

In this section, we will discuss the bifurcation structure of the system, the relation between the fluctuation and the energy landscape, and effects of the multiple-swallowtail structure on the superfluidity based on the results presented in Secs.~\ref{sec:Results} and \ref{sec:Results_unstable}.

From the results for the stable and unstable stationary solutions and their excitation spectra, we can discuss the bifurcation structures of the system. As shown in Figs.~\ref{fig:energy_diagram_sw_32_5_25} and \ref{fig:energy_diagram_sw_48_5_25}, the stable and the unstable branches merge at the critical velocity. It implies that a saddle-node (SN) type bifurcation, {\it i.e.}, either conventional SN or Hamiltonian SN, occurs in this system. The energy diagram and the scaling law shown in Sec.~\ref{subsec:excitation_spectra_2D} are consistent with the Hamiltonian saddle node (HSN) bifurcation \cite{Huepe2000,Pham2002}. 
The normal form of the HSN bifurcation is given by
\begin{eqnarray}
\frac{d^2}{d t^2}u(t)=\lambda-\beta u(t)^2,\label{eq:Results_normal_form_HSN}
\end{eqnarray}
where $u(t)$ is the amplitude of the critical mode, $t$ is the time, $\lambda\propto 1-v/v_{\rm c}$ denotes a bifurcation parameter, and $\beta$ is a constant related to the system parameters. The linear stability analysis around the stationary solution of Eq.~(\ref{eq:Results_normal_form_HSN}) ($u(t)=\sqrt{\lambda}$) shows that the frequency is proportional to $\lambda^{1/4}\propto (1-v/v_{\rm c})^{1/4}$. This is the same behavior of our systems. Physically, the normal form (\ref{eq:Results_normal_form_HSN}) shows that the breakdown of the metastable state is caused by the disappearance of the energy barrier, as described in Fig.~\ref{fig:schematic_picture}. 

We note that a similar mechanism for the destabilization of the metastable state has been found in attractive BECs in harmonic traps \cite{Ueda1998}. The attractive BEC collapses when the number of atoms in the harmonic trap exceeds a critical value. Near the collapse point, the monopole mode, which is a low-lying excitation in the attractive BEC, obeys a one-fourth power law, as found by variational calculations \cite{Ueda1998} and numerical calculations \cite{Huepe2003}. In Ref.~\cite{Huepe2003}, the HSN bifurcation was reported to occur in this system. We anticipate that the origin of the scaling law is the same as that for our system.

However, there is some evidence for the conventional SN bifurcation near the critical velocity of a superfluid in the presence of an obstacle \cite{Huepe2000,Pham2002,Kato2010,Watabe2013}. The time scale characteristic to the conventional SN bifurcation is proportional to the square root of the distance from the bifurcation points. In future work, we will seek to determine under what conditions the bifurcation near the critical velocity of a superfluid is a conventional SN or a Hamiltonian SN. Another area of future work is to derive the normal form given by Eq. (\ref{eq:Results_normal_form_HSN}), starting from the GP and Bogoliubov equations. 

We can also discuss the relation between the fluctuations and the energy landscape. The dynamics near the critical velocity are often discussed in terms of the energy landscape, which is schematically depicted in Fig.~\ref{fig:schematic_picture}.  Near the critical velocity, the metastable and unstable states approach each other in the configuration space, and thus the landscape around the local minimum becomes flat along a particular direction in the coordinate space. This results in the gap closing in the excitation spectrum, as shown in Sec. III C. The meaning of the abscissa in the energy landscape ({\it i.e.} coordinate of configuration space) is not generally clear except for a few cases where it was associated with collective coordinates \cite{Mueller2002, Eckel2014}. In the present case, we can identify the abscissa with the amplitude of the Bogoliubov eigenstate that has the lowest excitation energy. On the basis of the results in Sec.~\ref{subsec:fluctuation_2D}, we see that the motion of $\Psi$ from the metastable state to the unstable state is accompanied by a density fluctuation near the defect potential. 

The decay of supercurrent is frequently understood on the basis of the energy landscape as shown in Fig.~\ref{fig:schematic_picture}. 
Recall that the phase slip rate for one-dimensional ring systems has been calculated in \cite{McCumber1970,Polkovnikov2005}, where the conventional swallowtail structure plays a crucial role.

Our results imply that the conventional energy landscape, {\it i.e.,} the two local minima are separated by one local maximum, is valid only in the vicinity of the critical velocity (see Figs.~\ref{fig:energy_diagram_sw_32_5_25} and \ref{fig:energy_diagram_sw_48_5_25}). Away from the critical velocity, the energy landscape can not be written in the form shown in Fig.~\ref{fig:schematic_picture} and we expect that the multiple-swallowtail structure affects the phase slip rate. 

%%%%%%%%%%%%%%%%%%%%%%%%%%%%%%%%%%%%%%%%%%%%%%%%%%%%%%%%%%%%%%
\section{Summary}\label{sec:Summary}
%%%%%%%%%%%%%%%%%%%%%%%%%%%%%%%%%%%%%%%%%%%%%%%%%%%%%%%%%%%%%%

In summary, we investigated the metastability, excitations, fluctuations, and swallowtail structures of the BEC with a uniformly moving defect in a two-dimensional torus system. We first calculated the total energy and the total momentum as functions of the driving velocity of the moving defect. A negative effective-mass region appears near the critical velocity, as is the case for optical lattice systems. In contrast to optical lattice systems, the negative effective-mass states are metastable. This difference comes from that the DI in the optical lattice systems is due to the formation of the long-period structures, such as the period-doubling states and the bright gap solitons, which are prohibited in the torus systems. We also found GVPs in stationary states in the presence of a strong defect.

Using the results of the GP equation, we solved the Bogoliubov equation and obtained the excitation spectra. We determined that near the critical velocity, the scaling of the energy gap followed a one-fourth power law. This implies an algebraic divergence of the characteristic time scale toward the critical velocity and a violation of the adiabaticity condition at the critical velocity. 

From wave functions of the excited states $u_i(\bm{r})$ and $v_i(\bm{r})$, we obtained the fluctuation properties and showed that the density (amplitude of the order parameter) fluctuations are enhanced near the critical velocity. 

We also calculated the QD and found that it increased near the critical velocity. We confirmed the validity of the GP and the Bogoliubov approximations on the basis of these results.

We found unconventional swallowtail structures (multiple-swallowtail structure)  by the direct calculations of the unstable solutions. We discussed that the number of unstable branches depends on the width of the superflow path and expect that the multiple-swallowtail structure reduces to the conventional one in the narrow superflow path limit. 

We discussed the bifurcation structure of the system. The results for the one-fourth power-law scaling near the critical velocity and the unstable branches imply that the HSN bifurcation occurs in the system, which describes the disappearance of the energy barrier that protects a metastable state. We pointed out that the same scaling law appears in the attractive BEC in a harmonic trap near the collapse point. We also discussed the effects of the multiple-swallowtail structure on the calculations for the phase slip rate.

In future work, we will attempt to determine why the Hamiltonian saddle-node bifurcation appears at the critical velocity and to derive the normal form from the GP and Bogoliubov equations. The energy landscape away from the critical velocity remains open. Full knowledge of the energy landscape of the multiple-swallowtail structures will serve as the understanding of the breakdown of the superfluidity due to the vortex nucleations.

A possible further extension of the present work is to study the effects of quantum fluctuations on the metastability of superfluidity. In particular, these effects are crucial for cases that are not weakly interacting and that are near the critical velocity. In fact, a nonzero drag force acting on a defect below the critical velocity in a one-dimensional system has been reported in Ref.~\cite{Sykes2009}. This phenomenon is due to quantum fluctuations. It is important to understand the effects of quantum fluctuations on vortex nucleation.

Another extension of the present work is to study multicomponent systems, such as spinor BECs \cite{Kawaguchi2012,Stamper-Kurn2013}, and to reveal the effects of the internal degrees of freedom on the metastability of the superfluidity.

%%%%%%%%%%%%%%%%%%%%%%%%%%%%%%%%%%%%%%%%%%%%%%%%%%%%%%%%%%%%%%
%{\it Acknowledgment}\\
\begin{acknowledgments}
We thank S. Watabe and I. Danshita for fruitful discussions and D. Yamamoto for useful comments. We also thank an anonymous referee for his or her indication of the unstable solutions, which trigger the findings of the multiple-swallowtail structures. M. K. acknowledges the support of a Grant-in-Aid for JSPS Fellows (Grant No. 239376). This work is supported by JPSJ KAKENHI (Grant No. 24543061).
\end{acknowledgments}

%%%%%%%%%%%%%%%%%%%%%%%%%%%%%%%%%%%%%%%%%%%%%%%%%%%%%%%%%%%%%%
\appendix
%%%%%%%%%%%%%%%%%%%%%%%%%%%%%%%%%%%%%%%%%%%%%%%%%%%%%%%%%%%%%%
\section{Methods of Numerical Calculations}\label{app:Methods_of_numerical_calculations}
%%%%%%%%%%%%%%%%%%%%%%%%%%%%%%%%%%%%%%%%%%%%%%%%%%%%%%%%%%%%%%

In this appendix, we explain the methods used for the numerical calculations. Similar methods were used in our previous work \cite{Kunimi2012}.

In order to obtain the solutions of the time-independent GP equation (\ref{eq:Model_time-independent_GP_equation_in_mov_frame}), we used imaginary time propagation. The imaginary-time GP equation is given by
\begin{align}
-\hbar\frac{\partial}{\partial t}\Psi(\bm{r}, t)&=-\frac{\hbar^2}{2m}\nabla^2\Psi(\bm{r}, t)+U(\bm{r})\Psi(\bm{r})\nonumber \\
&-\mu(t)\Psi(\bm{r}, t)+g|\Psi(\bm{r}, t)|^2\Psi(\bm{r}, t),\label{eq:Appendix_imaginary_time_GP}
\end{align}
where $\mu(t)$ is the time-dependent chemical potential, whose time dependence is determined by the total number of particles (\ref{eq:Model_total_particle_number_condtion}). The time-independent solution of Eq.~(\ref{eq:Appendix_imaginary_time_GP}) coincides with the original GP equation (\ref{eq:Model_time-independent_GP_equation_in_mov_frame}).

From the twisted periodic boundary conditions (\ref{eq:Model_twisted_periodic_boundary_condtion_x}) and (\ref{eq:Model_twisted_periodic_boundary_condtion_y}), we can expand the condensate wave function as a series of plane waves, as follows:
\begin{eqnarray}
\Psi(\bm{r}, t)&=&\sqrt{n_0}\sum_{\bm{G}}C_{\bm{q}+\bm{G}}(t)e^{i(\bm{q}+\bm{G})\cdot\bm{r}},\label{eq:Appendix_expansion_of_the_condensate_wave_function}\\
\bm{G}&\equiv &\frac{2\pi}{L}(n_1\bm{e}_x+n_2\bm{e}_y),\label{eq:Appendix_definition_of_G}
\end{eqnarray}
where $C_{\bm{q}+\bm{G}}(t)$ is an expansion coefficient, $\bm{q}\equiv m\bm{v}/\hbar$, and $n_1$ and $n_2\in \mathbb{Z}$. Substituting Eq.~(\ref{eq:Appendix_expansion_of_the_condensate_wave_function}) into Eq.~(\ref{eq:Appendix_imaginary_time_GP}) and using the orthonormal condition $\int d\bm{r}e^{i(\bm{G}-\bm{G}')\cdot\bm{r}}=S\delta_{\bm{G}, \bm{G}'}$, we obtain the imaginary time GP equation for $C_{\bm{q}+\bm{G}}(t)$:
\begin{align}
-\hbar\frac{\partial}{\partial t}C_{\bm{q}+\bm{G}}(t)=\left[\frac{\hbar^2}{2m}(\bm{q}+\bm{G})^2-\mu(t)\right]C_{\bm{q}+\bm{G}}(t)\nonumber \\
+\frac{1}{S}\sum_{\bm{G}'}\bar{U}(\bm{G}-\bm{G}')C_{\bm{q}+\bm{G}'}(t)\nonumber\\
+g n_0\sum_{\bm{G},\Delta\bm{G}}C^{\ast}_{\bm{q}+\bm{G}'+\Delta\bm{G}}(t)C_{\bm{q}+\bm{G}'}(t)C_{\bm{q}+\bm{G}+\Delta\bm{G}}(t),\label{eq:Appendix_imaginary_time_GP_for_coefficients}
\end{align}
where $\bar{U}(\bm{k})$ is the Fourier transformation of the external potential:
\begin{eqnarray}
\bar{U}(\bm{k})\equiv \int d\bm{r}e^{-i\bm{k}\cdot\bm{r}}U(\bm{r}).\label{eq:Appendix_Fourier_transformation_external_potential}
\end{eqnarray}
The total particle number condition for $C_{\bm{q}+\bm{G}}(t)$ is given by
\begin{eqnarray}
1=\sum_{\bm{G}}|C_{\bm{q}+\bm{G}}(t)|^2.\label{eq:Appendix_total_particle_number_condition}
\end{eqnarray}

From the boundary conditions (\ref{eq:Model_boundary_condtion_for_u_x}), (\ref{eq:Model_boundary_condtion_for_v_x}), (\ref{eq:Model_boundary_condtion_for_u_y}), and (\ref{eq:Model_boundary_condtion_for_v_y}), the wave functions $u_i(\bm{r})$ and $v_i(\bm{r})$ can be expanded as a series of plane waves, as follows: 
\begin{eqnarray}
u_i(\bm{r})&=&\frac{1}{\sqrt{S}}\sum_{\bm{G}}A_{\bm{q}+\bm{G},i}e^{+i(\bm{q}+\bm{G})\cdot\bm{r}},\label{eq:Appendix_expansion_for_u}\\
v_i(\bm{r})&=&\frac{1}{\sqrt{S}}\sum_{\bm{G}}B_{\bm{q}+\bm{G},i}e^{-i(\bm{q}+\bm{G})\cdot\bm{r}},\label{eq:Appendix_expansion_for_v}
\end{eqnarray}
where $A_{\bm{q}+\bm{G}, i}$ and $B_{\bm{q}+\bm{G}, i}$ are the expansion coefficients of mode $i$. The normalization condition for the wave functions of the excited states becomes
\begin{eqnarray}
\sum_{\bm{G}}\left[|A_{\bm{q}+\bm{G}, i}|^2-|B_{\bm{q}+\bm{G}, i}|^2\right]=1.\label{eq:Appendix_normalization_condition_u_and_v}
\end{eqnarray}
Substituting Eqs.~(\ref{eq:Appendix_expansion_of_the_condensate_wave_function}), (\ref{eq:Appendix_expansion_for_u}), and (\ref{eq:Appendix_expansion_for_v}) into Eq.~(\ref{eq:Model_Bogoliubov_equation}), we obtain the Bogoliubov equation for the expansion coefficients:
\begin{align}
&D_{\bm{G}}A_{\bm{q}+\bm{G},i}+\frac{1}{S}\sum_{\bm{G}'}\bar{U}(\bm{G}-\bm{G}')A_{\bm{q}+\bm{G}', i}\nonumber \\
&+2g n_0\sum_{\bm{G}'}S_{\bm{G}, \bm{G}'}A_{\bm{q}+\bm{G}',i}\nonumber \\
&\quad -g n_0\sum_{\bm{G}'}W_{\bm{G}, \bm{G}'}B_{\bm{q}+\bm{G}',i}=\epsilon_iA_{\bm{q}+\bm{G}, i},\label{eq:Appendix_Bogoliubov_equation_for_A}\\
-&D_{\bm{G}}B_{\bm{q}+\bm{G},i}-\frac{1}{S}\sum_{\bm{G}'}\bar{U}^{\ast}(\bm{G}-\bm{G}')B_{\bm{q}+\bm{G}', i}\nonumber \\
&-2g n_0\sum_{\bm{G}'}S^{\ast}_{\bm{G}, \bm{G}'}B_{\bm{q}+\bm{G}',i}\nonumber \\
&\quad +g n_0\sum_{\bm{G}'}W^{\ast}_{\bm{G}, \bm{G}'}A_{\bm{q}+\bm{G}',i}=\epsilon_iB_{\bm{q}+\bm{G}, i},\label{eq:Appendix_Bogoliubov_equation_for_B}
\end{align}
where we have introduced the following variables in order to simplify the notation:
\begin{eqnarray}
D_{\bm{G}}&\equiv &\frac{\hbar^2}{2m}(\bm{q}+\bm{G})^2-\mu,\label{eq:Appendix_definition_of_D}\\
S_{\bm{G}, \bm{G}'}&\equiv &\sum_{\bm{G}''}C^{\ast}_{\bm{q}+\bm{G}''+\bm{G}'-\bm{G}}C_{\bm{q}+\bm{G}''},\label{eq:Appendix_definition_of_S}\\
W_{\bm{G}, \bm{G}'}&\equiv &\sum_{\bm{G}''}C_{\bm{q}+\bm{G}-\bm{G}''+\bm{G}'}C_{\bm{q}+\bm{G}''}.\label{eq:Appendix_definition_of_W}
\end{eqnarray}
Numerically diagonalizing Eqs.~(\ref{eq:Appendix_Bogoliubov_equation_for_A}) and (\ref{eq:Appendix_Bogoliubov_equation_for_B}), we obtain the excitation spectra and the wave functions of the excited states.

We introduced the cutoff $\bm{G}_{\rm c}$ to calculate the summation of $\bm{G}$. We used the cutoff wave number (the number of bases) $G_{\rm c}\xi=7.82(4973), 10.1(8227)$, and $11.4(10557)$ for $L=32\xi$ and $6.71(8227), 7.59(10557)$, and $8.38(12893)$ for $L=48\xi$. We checked that the cutoff dependence of the present results is negligibly small, other than for the calculation of the quantum depletion (see Ref. \cite{Note_depletion}).

%%%%%%%%%%%%%%%%%%%%%%%%%%%%%%%%%%%%%%%%%%%%%%%%%%%%%%%%%%%%%%
\section{Pseudo-Arclength Continuation Method}\label{app:PACM}
%%%%%%%%%%%%%%%%%%%%%%%%%%%%%%%%%%%%%%%%%%%%%%%%%%%%%%%%%%%%%%

In this appendix, we explain the pseudo-arclength continuation method (PACM) \cite{Keller1986}. We used this method to obtain the unstable solutions. This method was applied to the spin-1 GP equation in Ref.~\cite{Chen2011,Chen2014}.

We consider nonlinear algebraic equations,
\begin{eqnarray}
G_i(\bm{u}, \lambda)=0, \quad (i=1,\cdots, M),\label{eq:Appendix_equation_we_want_to_solve}
\end{eqnarray}
where $\bm{u}\in\mathbb{R}^M$ and $\lambda\in\mathbb{R}$ is a parameter. One way to solve this equation is the Newton method:
\begin{eqnarray}
\sum_{j=1}^M\left.\frac{\partial G_i(\bm{u}, \lambda)}{\partial u_j}\right|_{\bm{u}=\bm{u}_0}\delta u_j&=&-G_i(\bm{u}_0,\lambda),\label{eq:Appendix_Newton_method_applying_to_G}\\
\bm{u}_1&=&\bm{u}_0+\delta\bm{u},\label{eq:appendix_next_step_for_u}
\end{eqnarray}
where $\bm{u}_0$ and $\bm{u}_1$ are the initial and next values of $\bm{u}$, respectively. Solving Eq.~(\ref{eq:Appendix_Newton_method_applying_to_G}) until convergence, we can obtain a solution for Eq.~(\ref{eq:Appendix_equation_we_want_to_solve}). However, this method fails if the saddle-node bifurcation point exists. To avoid this difficulty, we use the PCAM method.

\begin{figure}[t]
\centering
\includegraphics[width=7.0cm,clip]{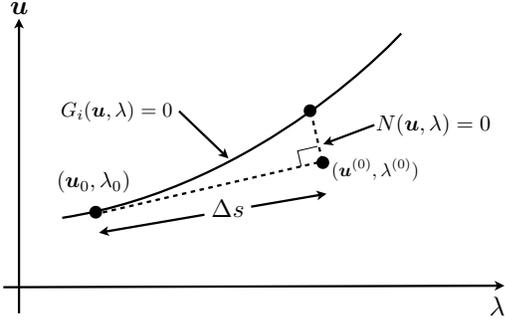}
\caption{(Color online) Schematic picture for the PCAM.}
\label{fig:schematic_picture_PCAM}
\end{figure}%

In the PCAM method, instead of solving Eq.~(\ref{eq:Appendix_equation_we_want_to_solve}) for fixed $\lambda$, we regard $\lambda$ as a variable and solve the following equations:
\begin{align}
G_i(\bm{u},\lambda)&=0,\quad (i=1,\cdots,M)\label{eq:Appendix_equation_for_PCAM_1}\\
N(\bm{u},\lambda)&=0,\label{eq:Appendix_equation_for_PCAM}\\
N(\bm{u},\lambda)&\equiv \dot{\bm{u}}_0\cdot(\bm{u}-\bm{u}_0)+\dot{\lambda}_0(\lambda-\lambda_0)-\Delta s,\label{eq:Appendix_definition_of_N}
\end{align}
where $\bm{u}_0$ is a solution of Eq.~(\ref{eq:Appendix_equation_we_want_to_solve}) for $\lambda=\lambda_0$, $\dot{\bm{u}}_0$ and $\dot{\lambda}_0$ are the normalized tangent vector in the $(\bm{u}, \lambda)$ space at the point $(\bm{u}_0, \lambda_0)$, and $\Delta s$ is the arclength (see Fig.~\ref{fig:schematic_picture_PCAM}). Equation $N(\bm{u}, \lambda)$=0 represents the plane that is perpendicular to the tangent vector $(\dot{\bm{u}}_0, \dot{\lambda}_0)$ and is separated by the distance $\Delta s$ from the point $(\bm{u}_0, \lambda_0)$. The tangent vector $(\dot{\bm{u}}_0, \dot{\lambda}_0)$ is determined by the following way: Let $s$ be a parameter that assigns the position of the orbit in the space $(\bm{u}, \lambda)$. The tangent vector can be obtained by solving the following equation:
\begin{eqnarray}
\left.\frac{d}{d s}G_i(\bm{u}(s), \lambda(s))\right|_{s=s_0}=0,\label{eq:Appendix_derivative_of_G_with_respective_to_s}
\end{eqnarray}
where $s_0$ is the value of $s$ at the point $(\bm{u}_0,\lambda_0)$. Equation (\ref{eq:Appendix_derivative_of_G_with_respective_to_s}) reduces to
\begin{eqnarray}
&&\sum_{j=1}^M\left.\frac{\partial G_i(\bm{u},\lambda)}{\partial u_j}\right|_{\bm{u}=\bm{u}_0,\lambda=\lambda_0}\dot{u}_{0,j}\nonumber \\
&&\hspace{6.0em}+\left.\frac{\partial G_i(\bm{u},\lambda)}{\partial\lambda}\right|_{\bm{u}=\bm{u}_0,\lambda=\lambda_0}\dot{\lambda}_0=0,\label{eq:Appendix_determine_tangent_vector}\\
&&\dot{u}_{0,j}\equiv\left.\frac{\partial u_{0,j}(s)}{\partial s}\right|_{s=s_0},\quad \dot{\lambda}_0\equiv \left.\frac{\partial \lambda(s)}{\partial s}\right|_{s=s_0}.\label{eq:Appendix_definition_dot_u_and_dot_lambda}
\end{eqnarray}
The normalization condition for the tangent vector is given by
\begin{eqnarray}
\dot{\bm{u}}^2+\dot{\lambda}^2=1.\label{eq:Appendix_normalization_condition}
\end{eqnarray}
From the normalization condition, the overall sign of the tangent vector is not determined. The overall sign can be determined so that the following condition is satisfied :
\begin{eqnarray}
\dot{\bm{u}}_{\rm p}\cdot\dot{\bm{u}}_0+\dot{\lambda}_{\rm p}\dot{\lambda}_0>0,\label{eq:Appendix_overall_factor}
\end{eqnarray}
where $(\dot{\bm{u}}_{\rm p}, \dot{\lambda}_{\rm p})$ is the tangent vector of the previous step.

We summarize the procedure for the PACM as follows:
\begin{itemize}
\item[(i)] Prepare $(\bm{u}_0, \lambda_0)$.
\item[(ii)] Set the appropriate value of $\Delta s$.
\item[(iii)] Calculate the tangent vector $(\dot{\bm{u}}_0, \dot{\lambda}_0)$ from Eqs.~(\ref{eq:Appendix_determine_tangent_vector}), (\ref{eq:Appendix_normalization_condition}), and (\ref{eq:Appendix_overall_factor}).
\item[(iv)] Calculate $(\bm{u}^{(0)}, \lambda^{(0)})\equiv (\bm{u}_0+\Delta s \dot{\bm{u}}_0, \lambda_0+\Delta s\dot{\lambda}_0)$ (Euler predictor). This is the initial condition of the Newton method (see Fig.~\ref{fig:schematic_picture_PCAM}).
\item[(v)] Solve Eq.~(\ref{eq:Appendix_equation_for_PCAM}) by the Newton method :
\begin{align}
&\sum_{j=1}^M\left.\frac{\partial G_i(\bm{u}, \lambda)}{\partial u_j}\right|_{\bm{u}=\bm{u}^{(0)},\lambda=\lambda^{(0)}}\delta u_j\nonumber \\
&+\left.\frac{\partial G_i(\bm{u},\lambda)}{\partial \lambda}\right|_{\bm{u}=\bm{u}^{(0)},\lambda=\lambda^{(0)}}\delta\lambda=-G_i(\bm{u}^{(0)},\lambda^{(0)}),\label{eq:Appendix_Newton_for_G}\\
&\sum_{j=1}^M\dot{u}_{0,j}\delta u_j+\dot{\lambda}_0\delta\lambda=-N(\bm{u}^{(0)}, \lambda^{(0)}),\label{eq:Appendix_Newton_for_N}\\
&\bm{u}^{(1)}=\bm{u}^{(0)}+\delta\bm{u},\quad \lambda^{(1)}=\lambda^{(0)}+\delta\lambda.\label{eq:Appendix_Newton_next_step}
\end{align}
\item[(iv)] Iterate (v) until convergence.
\end{itemize}

\newpage

\begin{thebibliography}{99}
\bibitem{note_fraction}
To obtain this value, we use a linear fit to more than $10$ data points of the total momentum for $v<0.05v_0$.
\bibitem{Note_depletion}
We use extrapolation to evaluate the quantum depletion. The method is as follows: In uniform two-dimensional systems, the cutoff dependence of the quantum depletion is given by $O(G_{\rm c}^{-2})$, where $G_{\rm c}$ is the cutoff wave number. Therefore, we assume the same cutoff dependence of the quantum depletion even in the presence of the defect potential. By adjusting $a$ and $b$, we can fit the function $N_{\rm dep}/N=a-b/(G_{\rm c}\xi)^2$ to three data points, and in this way, we can calculate the quantum depletion. The error bars for the fitting are not shown in Figs.~\ref{fig:depletion_32_and_48} and \ref{fig:dep_different_delta_48_1_25} because these are sufficiently small in the range of our graph.
\end{thebibliography}

%%%%%%%%%%%%%%%%%%%%%%%%%%%%%%%%%%%%%%%%%%%%%%%%%%%%%%%%%%%%%

\end{document}